\documentclass[prd,showpacs,preprintnumbers,twocolumn,amsmath,nofootinbib,amssymb]{revtex4}
\usepackage{graphicx,color,dcolumn,booktabs,bm}
\usepackage{longtable,lscape}
\usepackage{txfonts}
\usepackage{overpic}
\usepackage{amssymb}
\usepackage{epstopdf}
\usepackage{indentfirst}
\usepackage{feynmf}   %{feynmp}
\usepackage{slashed}  %for Feynman symbols
\usepackage{cases}
\usepackage{color}
\usepackage{float}
\usepackage{multirow}
\usepackage{graphicx,color,dcolumn,booktabs,bm}
\usepackage{epsfig,dsfont,amssymb,amsmath,amsfonts,amsbsy,mathrsfs}
\graphicspath{{Figures/}} %

\usepackage{hyperref}
\hypersetup{colorlinks,citecolor=blue,anchorcolor=red,menucolor=red, linkcolor=red,filecolor=red,runcolor=red,urlcolor=blue,frenchlinks=red}

\makeatletter
\@addtoreset{equation}{section}
\makeatother

\allowdisplaybreaks

\begin{document}

\title{Possible triple-charm molecular pentaquarks from $\Xi_{cc}D_1/\Xi_{cc}D_2^*$ interactions}

\author{Fu-Lai Wang$^{1,2}$}
\email{wangfl2016@lzu.edu.cn}
\author{Rui Chen$^{1,2}$}
\email{chenr15@lzu.edu.cn}
\author{Zhan-Wei Liu$^{1,2}$}
\email{liuzhanwei@lzu.edu.cn}
\author{Xiang Liu$^{1,2}$}
\email{xiangliu@lzu.edu.cn}
\affiliation{
$^1$School of Physical Science and Technology, Lanzhou University, Lanzhou 730000, China\\
$^2$Research Center for Hadron and CSR Physics, Lanzhou University
and Institute of Modern Physics of CAS, Lanzhou 730000, China
}

\begin{abstract}
In this work, we explore a systematic investigation on $S$-wave interactions  between a doubly charmed baryon $\Xi_{cc}(3621)$ and a charmed meson in a $T$ doublet  $(D_1,\,D_2^*)$. We first analyze the possibility for forming $\Xi_{cc}D_1/\Xi_{cc}D_2^*$ bound states with the heavy quark spin symmetry. Then, we further perform a dynamical study on the $\Xi_{cc}D_1/\Xi_{cc}D_2^*$ interactions within a one-boson-exchange model by considering both the $S$-$D$ wave mixing and coupled channel effect. Finally, our numerical results conform the proposals from the heavy quark spin symmetry analysis: the $\Xi_{cc}D_1$ systems with $I(J^P)=$ $0(1/2^+,\,3/2^+)$ and the $\Xi_{cc}D_2^*$ systems with $I(J^P)=$ $0(3/2^+,\,5/2^+)$ can possibly be loose triple-charm molecular pentaquarks. Meanwhile, we also extend our model to the $\Xi_{cc}\bar{D}_1$ and $\Xi_{cc}\bar D_2^*$ systems, and our results indicate the isoscalars of $\Xi_{cc}\bar{D}_1$ and $\Xi_{cc}\bar{D}_2^*$ can be possible molecular candidates.
\end{abstract}

\pacs{12.39.Pn, 14.40.Lb, 14.20.Lq}

\maketitle

\section{introduction}\label{sec1}

Since 2003, people have paid more and more attention to the study of exotic states that are very different from conventional mesons and baryons (made up of a pair of quark and antiquark, and three quarks, respectively). The study of exotic states can deepen our understanding of the nonperturbative behavior of quantum chromodynamics (QCD).

In particular, inspired by the near threshold observations of $X/Y/Z$ and $P_c$ states (see Refs. \cite{Chen:2016qju,Liu:2013waa,Hosaka:2016pey} for review), hadron-hadron interactions are studied in a heavy flavor sector, through which the inner structures and underlying mechanism of new discoveries could be explored. Meanwhile, other possible hidden-charm tetraquark and pentaquark exotic molecules are proposed and studied.

Last year, the LHCb Collaboration reported an important observation of a doubly charmed baryon $\Xi_{cc}(3621)$ in the $\Lambda_c^+K^-\pi^+\pi^+$ invariant mass spectrum \cite{Aaij:2017ueg}. This discovery not only complements the baryon family, but also provides us a good opportunity to study the interactions involved in a doubly charmed baryon. Moreover, we can search for possible double-charm, triple-charm, and tetrad-charm multiquark molecules by constructing a doubly charmed baryon and a nucleon system \cite{Meng:2017udf},  a doubly charmed baryon and a charmed baryon system \cite{Chen:2018pzd}, a doubly charmed baryon and a doubly charmed baryon system \cite{Meng:2017fwb}, and a doubly charmed baryon and a charmed meson system \cite{Chen:2017jjn}. Here, we also notice another theoretical prediction of triple-charm  pentaquarks \cite{Azizi:2018dva,Guo:2013xga}.

As discussed in Ref.\cite{Chen:2017jjn}, if the near threshold structures, the $X/Y/Z$ and $P_c$ states, can be , respectively, identified as hidden-charm tetraquark and pentaquark molecules, one can also propose that there may exist possible triple-charm pentaquark molecular states, which are composed by a doubly charmed baryon and a charmed meson. In Ref. \cite{Chen:2017jjn}, after checking the $\Xi_{cc}D/\Xi_{cc}D^*$ interactions, a $\Xi_{cc}D$ state with $I(J^P)=0(1/2^-)$ and a $\Xi_{cc}D^*$ state with $I(J^P)=0(3/2^-)$ can be recommended as possible triple-charm pentaquarks molecular candidates.

In a heavy quark limit, heavy mesons can be categorized into different doublets based on the heavy quark spin symmetry, i.e., $H=\left(D,D^*\right)$, $S=\left(D_0,D_1^{\prime}\right)$, $T=\left(D_1,D_2^*\right)$, which correspond to $j_l^P=1/2^-$, $1/2^+$, and $3/2^+$, respectively. In this work, we want to further study the interactions between a doubly charmed baryon $\Xi_{cc}(3621)$ and a charmed meson in a $T$ doublet, $D_1(2420)$ and $D_2^*(2460)$. Meanwhile, we discuss the properties of heavy quark spin symmetry in the interactions between a doubly charmed baryon and a charmed meson. The obtained information is valuable to predict possible triple-charm pentaquark molecules, the $\Xi_{cc}D_1$ and $\Xi_{cc}D_2^*$ systems.

In this paper, we also perform quantitatively calculations on the $\Xi_{cc}D_1/\Xi_{cc}D_2^*$ interactions by using a one-boson-exchange model (OBE), which is often adopted to study the heavy flavor hadron interactions and identify newly $X/Y/Z$ and $P_c$ states in a hadronic molecular picture. All of coupling constants are estimated with the help of nucleon couplings within a quark model. Both the $S$-$D$ wave mixing and coupled channel effect will be considered.

This paper is organized as follows. In Sec. \ref{sec2}, we give a heavy quark spin symmetry analysis of the $S$-wave interaction between a doubly charmed baryon and a charmed meson in the $T$ doublet. We study the effective potentials with dynamical effects considered by adopting the OBE model in Sec. \ref{sec3}, and the corresponding numerical results are presented in Sec. \ref{sec4}. We end with a conclusion and discussion in Sec. \ref{sec5}.

\section{Heavy quark spin symmetry analysis}\label{sec2}

According to heavy quark spin symmetry (HQSS), hadrons containing a single heavy quark with total spin $J_{\pm}=j_l\pm1/2$ (except $j_l=0$) come into doublets. Heavy hadrons in the same doublet are approximatively degenerate. In this work, we will study the $S$-wave interaction between a doubly charmed baryon $\Xi_{cc}(3621)$ and a charmed meson in the $T$ doublet. Here, the $T$ doublet includes $D_1$ and $D_2^*$ mesons with $J^P=1^+$ and $2^+$, respectively.

\begin{figure}[!htbp]
\center
\includegraphics[width=0.43\textwidth]{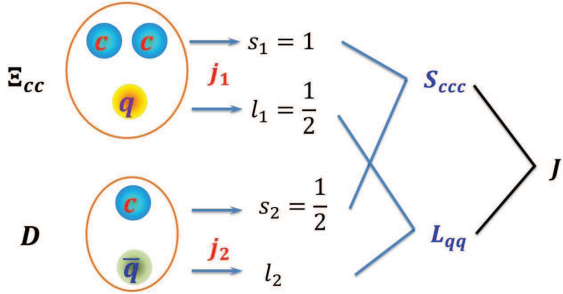}\\
\caption{Diagram for $9-j$ coefficients in a heavy quark basis. Here,  $j_1$ and $j_2$ are the spins for the doubly-charmed baryon and charmed meson, respectively. Inside the hadrons, $s_1$ and $s_2$ stand for the spins of the heavy quarks, while $l_1$ and $l_2$ are the total angular momentum for the light quarks. $l_2=3/2$ for $D_1/D_2^*$ mesons, and $l_2=1/2$ for $D/D^*$ mesons.}\label{HQSS}
\end{figure}

In general, the HQSS plays an important role for hadron-hadron interaction in heavy flavor sectors. In order to perform an HQSS analysis, we should first expand the spin wave functions of heavy hadron systems in terms of a heavy quark basis, i.e.,
\begin{widetext}
\begin{eqnarray}
\left|\ell_1 s_1 j_1; \ell_2 s_2 j_2; J M \right\rangle &=& \sum_{S,L}\left[(2S+1)(2L+1)(2j_1+1)(2j_2+1)\right]^{1/2} \left\{
\begin{array}{ccc}
\ell_1 & \ell_2 & L \\
s_1 & s_2 & S \\
j_1 & j_2 & J
\end{array}
\right\} \ |\ell_1 \ell_2 L; s_1 s_2 S; J M \big\rangle.
\end{eqnarray}
\end{widetext}
In Fig. \ref{HQSS}, we present the diagram for $9-j$ coefficients in the heavy quark basis, where the spins for heavy quarks and light quarks have been combined into $S_{ccc}$ and $L$, respectively. In the $m_Q\to\infty$ limit, $S_{ccc}$, $J$, and $L$ are conserved quantum numbers.

Once expanding the spin wave functions for the discussed systems, we can obtain
\begin{eqnarray}
|\Xi_{cc} D(J=\frac{1}{2})\big \rangle &=& \frac{1}{2}|S_{ccc}=\frac{1}{2},\, L=0; J=\frac{1}{2}\big \rangle_{\ell_2=\frac{1}{2}}\nonumber\\
&&+\frac{1}{2\sqrt{3}}|S_{ccc}=\frac{1}{2},\, L=1; J=\frac{1}{2}\big \rangle_{\ell_2=\frac{1}{2}}\nonumber\\
&&+\sqrt{\frac{2}{3}}|S_{ccc}=\frac{3}{2},\, L=1; J=\frac{1}{2}\big \rangle_{\ell_2=\frac{1}{2}},\nonumber\\\\
%%%%
|\Xi_{cc} D^{*}(J=\frac{3}{2})\big\rangle &=&-\frac{1}{3}|S_{ccc}=\frac{1}{2},\, L=1; J=\frac{3}{2}\big \rangle_{\ell_2=\frac{1}{2}}\nonumber\\
&&+\frac{1}{\sqrt{3}}|S_{ccc}=\frac{3}{2},\, L=0; J=\frac{3}{2}\big \rangle_{\ell_2=\frac{1}{2}}\nonumber\\
&&+\frac{\sqrt{5}}{3}|S_{ccc}=\frac{3}{2},\, L=1; J=\frac{3}{2}\big \rangle_{\ell_2=\frac{1}{2}},\nonumber\\\\
%%%%
|\Xi_{cc} D_1(J=\frac{1}{2})\big \rangle &=& -\frac{1}{3}|S_{ccc}=\frac{1}{2},\, L=1; J=\frac{1}{2}\big \rangle_{\ell_2=\frac{3}{2}}\nonumber\\
&&-\frac{1}{3\sqrt{2}}|S_{ccc}=\frac{3}{2},\, L=1; J=\frac{1}{2}\big \rangle_{\ell_2=\frac{3}{2}}\nonumber\\
&&-\sqrt{\frac{5}{6}}|S_{ccc}=\frac{3}{2},\, L=2; J=\frac{1}{2}\big \rangle_{\ell_2=\frac{3}{2}},\nonumber\\\\
%%%%
|\Xi_{cc} D_1(J=\frac{3}{2})\big \rangle &=&\frac{7}{12}|S_{ccc}=\frac{1}{2},\, L=1; J=\frac{3}{2}\big \rangle_{\ell_2=\frac{3}{2}}\nonumber\\
&&+\frac{1}{4}\sqrt{\frac{5}{3}}|S_{ccc}=\frac{1}{2},\, L=2; J=\frac{3}{2}\big \rangle_{\ell_2=\frac{3}{2}}\nonumber\\
&&+\frac{\sqrt{5}}{6}|S_{ccc}=\frac{3}{2},\, L=1; J=\frac{3}{2}\big \rangle_{\ell_2=\frac{3}{2}}\nonumber\\
&&+\frac{1}{2}\sqrt{\frac{5}{3}}|S_{ccc}=\frac{3}{2},\, L=2; J=\frac{3}{2}\big \rangle_{\ell_2=\frac{3}{2}},\nonumber\\\\
%%%%
|\Xi_{cc} D_2^{*}(J=\frac{3}{2})\big \rangle &=&\frac{1}{4}\sqrt{\frac{5}{3}}|S_{ccc}=\frac{1}{2},\, L=1;J=\frac{3}{2}\big\rangle_{\ell_2=\frac{3}{2}}\nonumber\\
&&+\frac{3}{4}|S_{ccc}=\frac{1}{2},\, L=2; J=\frac{3}{2}\big \rangle_{\ell_2=\frac{3}{2}}\nonumber\\
&&-\frac{1}{2\sqrt{3}}|S_{ccc}=\frac{3}{2},\, L=1; J=\frac{3}{2}\big \rangle_{\ell_2=\frac{3}{2}}\nonumber\\
&&-\frac{1}{2}|S_{ccc}=\frac{3}{2},\, L=2; J=\frac{3}{2}\big \rangle_{\ell_2=\frac{3}{2}},\nonumber\\\\
%%%%
|\Xi_{cc} D_2^{*}(J=\frac{5}{2})\big \rangle&=&-\frac{1}{3}|S_{ccc}=\frac{1}{2},\, L=2;J=\frac{5}{2}\big\rangle_{\ell_2=\frac{3}{2}}\nonumber\\
&&+\frac{1}{\sqrt{2}}|S_{ccc}=\frac{3}{2},\, L=1; J=\frac{5}{2}\big \rangle_{\ell_2=\frac{3}{2}}\nonumber\\
&&+\frac{1}{3}\sqrt{\frac{7}{3}}|S_{ccc}=\frac{3}{2},\, L=2; J=\frac{5}{2}\big \rangle_{\ell_2=\frac{3}{2}}.\nonumber\\
\end{eqnarray}

After adopting the above orthogonal HQSS basis, matrix elements for different interactions satisfy
\begin{eqnarray}\label{eqn1}
&&{}_{\ell_2'}\big\langle S'_{ccc},\,L'; J',\,\alpha'|H^{QCD}|S_{ccc},\, L; J,\, \alpha \big \rangle_{\ell_2}\nonumber\\
&=& \delta_{\alpha \alpha'}\delta_{JJ'}\delta_{S'_{ccc}S_{ccc}}\delta_{ L L'}\,\,\, {}_{\ell_2'}\big\langle L;\alpha||H^{QCD}||L; \alpha \big\rangle_{\ell_2}.
\end{eqnarray}
Here, $\alpha$ stands for other QCD conserved quantum numbers (like isospin and hypercharge, etc.). In $m_Q\to\infty$, transition matrix elements for QCD interactions between a different physical meson-baryon basis are related to the spin and flavor wave functions \cite{Garcia-Recio:2013gaa}. For $\Xi_{cc}D/\Xi_{cc}D^*$ and $\Xi_{cc}D_1/\Xi_{cc}D_2^*$ systems with given quantum number configurations $|S_{ccc}, L; J, \alpha\rangle$, one obtains
%\begin{widetext}
\begin{eqnarray}
&&{}_{\ell_2'=\frac{1}{2}}\big\langle L=0;\alpha||H^{QCD}||L=0; \alpha \big\rangle_{\ell_2=\frac{1}{2}}\nonumber\\
&\simeq&
{}_{\ell_2'=\frac{3}{2}}\big\langle L=2;\alpha||H^{QCD}||L=2; \alpha \big\rangle_{\ell_2=\frac{3}{2}},\\
&&{}_{\ell_2'=\frac{1}{2}}\big\langle L=1;\alpha||H^{QCD}||L=1; \alpha \big\rangle_{\ell_2=\frac{1}{2}}\nonumber\\
&\simeq&
{}_{\ell_2'=\frac{3}{2}}\big\langle L=1;\alpha||H^{QCD}||L=1; \alpha \big\rangle_{\ell_2=\frac{3}{2}}.
\end{eqnarray}
%\end{widetext}

Finally, we can find a serial of approximative relations between $\Xi_{cc}D_1/\Xi_{cc}D_2^*$ interactions and $\Xi_{cc}D/\Xi_{cc}D^*$ interactions,
\begin{eqnarray}
V_{\Xi_{cc}D_1}^{J=1/2} &=& -6V_{\Xi_{cc}D}^{J=1/2}+7V_{\Xi_{cc}D^*}^{J=3/2},\label{eqV12XiccD1}\\
V_{\Xi_{cc}D_1}^{J=3/2} &=& \frac{1}{4}\left(-9V_{\Xi_{cc}D}^{J=1/2}+13V_{\Xi_{cc}D^*}^{J=3/2}\right),\\
V_{\Xi_{cc}D_2^*}^{J=3/2} &=& \frac{1}{4}\left(-23V_{\Xi_{cc}D}^{J=1/2}+27V_{\Xi_{cc}D^*}^{J=3/2}\right),\\
V_{\Xi_{cc}D_2^*}^{J=5/2} &=& -2V_{\Xi_{cc}D}^{J=1/2}+3V_{\Xi_{cc}D^*}^{J=3/2}.
\end{eqnarray}

At present, we have already studied the $\Xi_{cc}D/\Xi_{cc}D^*$ interactions \cite{Chen:2017jjn}. After adopting the OBE model and considering the $S$-$D$ wave mixing effect, the $\Xi_{cc}D$ state with $I(J^P)=0(1/2^-)$ and the $\Xi_{cc}D^*$ state with $I(J^P)=0(3/2^-)$ are possible triple-charm pentaquark molecules, and the $\Xi_{cc}D^*$ state with $I(J^P)=0(3/2^-)$ is much stabler than the $\Xi_{cc}D$ state with $I(J^P)=0(1/2^-)$. Thus, we can conclude that the OBE effective potential for the $\Xi_{cc}D^*$ state with $I(J^P)=0(3/2^-)$ is attractive more strongly than that for the $\Xi_{cc}D$ state with $I(J^P)=0(1/2^-)$, i.e.,
\begin{eqnarray}
V^{I(J^P)=0(3/2^-)}_{\Xi_{cc}D^*}<V^{I(J^P)=0(1/2^-)}_{\Xi_{cc}D}<0. \label{eqxd}
\end{eqnarray}

Using Eqs. (\ref{eqV12XiccD1})$-$(\ref{eqxd}), we can estimate that
\begin{itemize}
  \item the interactions for the $\Xi_{cc}D_1$ system with $I(J^P)=0(1/2^+),\,0(3/2^+)$ and the $\Xi_{cc}D_2^*$ system with $I(J^P)=0(3/2^+),\,0(5/2^+)$ are all attractive.
  \item compared to the effective potentials for the $\Xi_{cc}D^*$ state with $I(J^P)=0(3/2^-)$, effective potentials for the $\Xi_{cc}D_1$ system with $I(J^P)=0(1/2^+),\,0(3/2^+)$ and $\Xi_{cc}D_2^*$ system with $I(J^P)=0(3/2^+),\,0(5/2^+)$ should be much stronger.
  \item since the interaction for the $\Xi_{cc}D^*$ state with $0(3/2^-)$ is attractive much more strongly than that for the $\Xi_{cc}D$ state with $0(1/2^-)$, we can obtain two relations: $V_{\Xi_{cc}D_1}^{I(J^P)=0(1/2^+)}<V_{\Xi_{cc}D_1}^{I(J^P)=0(3/2^+)}<0$, $V_{\Xi_{cc}D_2^*}^{I(J^P)=0(3/2^+)}<V_{\Xi_{cc}D_2^*}^{I(J^P)=0(5/2^+)}<0$.
  \item together with the weaker kinetic terms as their smaller reduced masses, the interactions for $\Xi_{cc}D_1$ systems with $I(J^P)=0(1/2^+),\,0(3/2^+)$ and $\Xi_{cc}D_2^*$ systems with $I(J^P)=0(3/2^+),\,0(5/2^+)$ can be strong enough to bind as bound states.
\end{itemize}

\section{Interactions within a dynamical study}\label{sec3}

After qualitatively analyzing the features of the $S$-wave interactions with HQSS between a doubly charmed baryon and a charmed meson in the $T$ doublet, here we adopt the OBE model to study the $\Xi_{cc}D_1(2420)$ and $\Xi_{cc}D_2^*(2460)$ interactions quantitatively. Meanwhile, both the $S$-$D$ wave mixing effect and coupled channel effect will be taken into consideration.

The flavor and spin-orbit wave functions for the discussed $\Xi_{cc}D_1$ and $\Xi_{cc}D_2^*$ systems are collected in Table \ref{flavor}.
\renewcommand\tabcolsep{0.15cm}
\renewcommand{\arraystretch}{1.8}
\begin{table}[!htpb]
\centering
\caption{The flavor and spin-orbit wave functions for the discussed $\Xi_{cc}D_1$ and $\Xi_{cc}D_2^*$ systems.}\label{flavor}
\begin{tabular}{ccccc}\toprule[1.0pt]\midrule[1.0pt]
     &$|I,I_3\rangle$    & Flavor    &$J^P$    &$|{}^{2S+1}L_J\rangle$\\\midrule[1.0pt]
$\Xi_{cc}D_{1}$&$|1,1\rangle$ & $\Xi_{cc}^{++}D_{1}^{+}$
&$1/2^+$    &$|{}^2\mathbb{S}_{\frac{1}{2}}/|{}^4\mathbb{D}_{\frac{1}{2}}\rangle$\\
&$|1,0\rangle$ & $\frac{1}{\sqrt{2}}(\Xi_{cc}^{++}D_{1}^{0}-\Xi_{cc}^{+}D_{1}^{+})$
&$3/2^+$    &$|{}^4\mathbb{S}_{\frac{3}{2}}/{}^2\mathbb{D}_{\frac{3}{2}}/{}^4\mathbb{D}_{\frac{3}{2}}\rangle$\\
               &$|1,-1\rangle$ & $\Xi_{cc}^{+}\,D_{1}^{0}$ \\
               &$|0,0\rangle$& $\frac{1}{\sqrt{2}}(\Xi_{cc}^{++}D_{1}^{0}+\Xi_{cc}^{+}D_{1}^{+})$ \\
$\Xi_{cc}D_{2}^{*}$ &$|1,1\rangle$ & $\Xi_{cc}^{++}D_{2}^{*+}$
&$3/2^+$    &$|{}^4\mathbb{S}_{\frac{3}{2}}/{}^4\mathbb{D}_{\frac{3}{2}}/{}^6\mathbb{D}_{\frac{3}{2}}\rangle$\\
                    &$|1,0\rangle$  & $\frac{1}{\sqrt{2}}(\Xi_{cc}^{++}D_{2}^{*0}-\Xi_{cc}^{+}D_{2}^{*+})$
&$5/2^+$  &$|{}^5\mathbb{S}_{\frac{5}{2}}/|{}^4\mathbb{D}_{\frac{5}{2}}/{}^6\mathbb{D}_{\frac{5}{2}}\rangle$\\
                    &$|1,-1\rangle$    & $\Xi_{cc}^{+}D_{2}^{*0}$ \\
                    &$|0,0\rangle$ & $\frac{1}{\sqrt{2}}(\Xi_{cc}^{++}D_{2}^{*0}+\Xi_{cc}^{+}D_{2}^{*+})$ \\
\bottomrule[1.0pt]\midrule[1.0pt]
\end{tabular}
\end{table}

Here, the general expressions for the spin-orbital wave functions $|{}^{2S+1}L_{J}\rangle$ are written as
\begin{eqnarray}
|\Xi_{cc}D_{1}({}^{2S+1}L_{J})\rangle &=&\sum_{m,m',m_Sm_L}C^{S,m_S}_{\frac{1}{2}m,1m'}
C^{J,M}_{Sm_S,Lm_L}
          \chi_{\frac{1}{2}m}\epsilon^{m'}|Y_{L,m_L}\rangle,\nonumber\\\\
|\Xi_{cc}D_{2}^{*}({}^{2S+1}L_{J})\rangle &=& \sum_{m,m'',m_Sm_L}C^{S,m_S}_{\frac{1}{2}m,2m''}
C^{J,M}_{Sm_S,Lm_L}
          \chi_{\frac{1}{2}m}\zeta^{m''}|Y_{L,m_L}\rangle.\nonumber\\
\end{eqnarray}
$C^{J,M}_{Sm_S,Lm_L}$, $C^{S,m_S}_{\frac{1}{2}m,1m'}$ and $C^{S,m_S}_{\frac{1}{2}m,2m''}$ are the Clebsch-Gordan coefficients, $\chi_{\frac{1}{2}m}$ and $Y_{L,m_L}$ correspond to the spin wave function and the spherical harmonics function, respectively. $\epsilon^{m'} (m'=0,\pm1)$ is defined as the polarization vector for $D_1$, with
\begin{eqnarray}
\epsilon^{\pm1}= \frac{1}{\sqrt{2}}\left(0,\pm1,i,0\right),\quad\quad
\epsilon^{0}= \left(0,0,0,-1\right).\nonumber
\end{eqnarray}
$\zeta^{m''}(m''=0,\pm1,\pm2)$ denotes the polarization tensor for $D_2^*$, which can be constructed by $\zeta^{m''}=\sum_{m_1,m_2}\langle1,m_1;1,m_2|2,m''\rangle\epsilon^{m_1}\epsilon^{m_2}$ \cite{Cheng:2010yd}.

\subsection{Effective Lagrangians}
According to the heavy quark symmetry, the chiral symmetry, and the hidden gauge symmetry \cite{Wise:1992hn,Casalbuoni:1992gi,Casalbuoni:1996pg,Yan:1992gz,Ding:2008gr}, the OBE effective Lagrangians for charmed mesons in the $T$ doublet and the light scalar, pesudoscalar and vector mesons are constructed as
\begin{eqnarray}
\label{eq:lag1}
\mathcal{L}&=&g''_{\sigma}\langle T^{(Q)\mu}_a\sigma\overline{T}^{\,({Q})}_{a\mu}\rangle+ik\langle T^{(Q)\mu}_{b}{\cal A}\!\!\!\slash_{ba}\gamma_5\overline{T}^{\,(Q)}_{a\mu}\rangle\nonumber\\
&&+i\beta^{\prime\prime}\langle T^{(Q)\lambda}_bv^{\mu}({\cal V}_{\mu}-\rho_{\mu})_{ba}\overline{T}^{\,(Q)}_{a\lambda}\rangle\nonumber\\
&&+i\lambda^{\prime\prime}\langle T^{(Q)\lambda}_b\sigma^{\mu\nu}F_{\mu\nu}(\rho)_{ba}\overline{T}^{\,(Q)}_{a\lambda}\rangle,
\end{eqnarray}
with the velocity $v=(1,\bm{0})$. Surperfield $T$ is defined as a linear combination of charmed mesons in $T$ doublets, i.e.,
\begin{eqnarray}
T_a &=& \frac{1+\slash \!\!\!v}{2}\left[D^{*\mu\nu}_{2a}\gamma_{\nu}-\sqrt{\frac{3}{2}}
D_{1a\nu}\gamma_5\left(g^{\mu\nu}-\frac{1}{3}\gamma^{\nu}(\gamma^{\mu}-v^{\mu})\right)\right].\quad\quad
\end{eqnarray}
Its conjugate field satisfies $\bar{T}_a=\gamma^0T_a^\dag\gamma^0$. The notation $\langle...\rangle$ stands for the trace of matrices in the spin and flavor space. Vector meson field $\rho_{\mu}$ and vector meson strength tensor $F_{\mu\nu}(\rho)$ are, respectively, defined as
\begin{eqnarray}
\rho_{\mu}&=&i\frac{g_V}{\sqrt{2}}\mathbb{V}_{\mu},\\
F_{\mu\nu}(\rho)&=&\partial_{\mu}\rho_{\nu}-\partial_{\nu}\rho_{\mu}+[\rho_{\mu},\rho_{\nu}].
\end{eqnarray}
In the above expressions, the vector current ${\cal V}_{\mu}$ and the axial current $\mathcal{A}_\mu$ are
\begin{eqnarray}
{\mathcal V}_{\mu}&=&\frac{1}{2}(\xi^{\dagger}\partial_{\mu}\xi+\xi\partial_{\mu}\xi^{\dagger}),\\
{\mathcal A}_{\mu}&=&\frac{1}{2}(\xi^\dag\partial_\mu\xi-\xi \partial_\mu\xi^\dag),
\end{eqnarray}
with $\xi=\exp(i\mathbb{P}/f_\pi)$ and the pion decay constant is taken as $f_\pi=132~\rm{MeV}$. At the leading order, the vector current and the axial current are
\begin{eqnarray}
\mathcal{V}_{\mu}&=&0,\\
\mathcal{A}_\mu&=&\frac{i}{f_\pi}\partial_\mu{\mathbb P},
\end{eqnarray}
respectively. Pseudoscalar meson matrices $\mathbb{P}$ and vector meson matrices $\mathbb{V}_{\mu}$ are expressed as
\begin{eqnarray*}
\mathbb{P}&=&\left(\begin{array}{cc}
\frac{\pi^{0}}{\sqrt{2}}+\frac{\eta}{\sqrt{6}}&\pi^{+}\\
\pi^{-}&-\frac{\pi^{0}}{\sqrt{2}}+\frac{\eta}{\sqrt{6}}\\
\end{array}\right),\\
\mathbb{V}_{\mu}&=&\left(\begin{array}{cc}
\frac{\rho^{0}}{\sqrt{2}}+\frac{\omega}{\sqrt{2}}&\rho^{+}\\
\rho^{-}&-\frac{\rho^{0}}{\sqrt{2}}+\frac{\omega}{\sqrt{2}}
\end{array}\right)_{\mu},\label{vector}
\end{eqnarray*}
respectively.

After expanding Eq. (\ref{eq:lag1}), we further obtain the concrete effective Lagrangians,
\begin{eqnarray}
\label{eq:lagp}
\mathcal{L}_{D_1 D_1\sigma} &=& -2g_\sigma^{\prime\prime}D_{1a\mu}D^{\mu\dagger}_{1a} \sigma ,\\
\mathcal{L}_{D^*_2D^*_2\sigma} &=& 2g_\sigma^{\prime\prime}D^{*\dagger}_{2a\mu\nu}D^{*\mu\nu}_{2a}\sigma ,\\
\mathcal{L}_{D_1D^*_2\sigma} &=& \sqrt{\frac{2}{3}}ig_\sigma^{\prime\prime}\epsilon^{\mu \nu \rho \tau}v_{\rho}(D^{\dagger}_{1a\nu}D^{*}_{2a\mu\tau}-D_{1a\nu}D^{*\dagger}_{2a\mu\tau}) \sigma ,\\
\mathcal {L}_{D_1 D_1\mathbb{P}}&=&-\frac{5ik}{3f_\pi}~\epsilon^{\mu\nu\rho\tau}v_\tau
    D^{\dagger}_{1a\mu}D_{1b\nu} \partial_\rho\mathbb{P}_{ba},\\
\mathcal {L}_{D^*_2D^*_2\mathbb{P}}&=&\frac{2ik}{f_\pi}~\epsilon^{\mu\nu\rho\tau}v_\nu
   D^{*\alpha\dagger}_{2a\rho}D^{*}_{2b\alpha\tau}\partial_\mu\mathbb{P}_{ba},\\
\mathcal {L}_{D_1D^*_2\mathbb{P}}&=&-\sqrt{\frac{2}{3}}\frac{k}{f_\pi}(D^{\dagger}_{1a\mu}
D^{*\mu\lambda}_{2b}+D_{1b\mu}D^{*\mu\lambda\dagger}_{2a})
\partial_\lambda\mathbb{P}_{ba},\\
\mathcal {L}_{D_1 D_1\mathbb{V}} &=& -\sqrt{2}\beta^{\prime \prime}
g_{V}(v\cdot\mathbb{V}_{ba}) D_{1b\mu}D^{\mu\dagger}_{1a}\nonumber\\
&&+\frac{5\sqrt{2}i\lambda^{\prime\prime} g_{V}}{3}(D^{\nu\dagger}_{1a}D^{\mu}_{1b}-D^{\nu}_{1b}D^{\mu\dagger}_{1a})\partial_\mu \mathbb{V}_{ba\nu},\\
\mathcal {L}_{D^*_2D^*_2\mathbb{V}} &=& \sqrt{2}\beta^{\prime \prime}
g_{V}(v\cdot\mathbb{V}_{ba}) D_{2b}^{*\lambda\nu}  D^{*\dagger}_{2a{\lambda\nu}}+2\sqrt{2}i\lambda^{\prime\prime} g_{V}\nonumber\\
&&(D^{*\lambda\nu}_{2b} D^{*\mu\dagger}_{2a\lambda}-D^{*\lambda\nu\dagger}_{2a}D^{*\mu}_{2b\lambda} )\partial_\mu \mathbb{V}_{ba\nu},\\
\mathcal {L}_{D_1D^*_2\mathbb{V}} &=& \frac{i\beta^{\prime \prime}g_{V}}{\sqrt{3}}\epsilon^{\lambda\alpha\rho\tau}v_{\rho}(v\cdot
\mathbb{V}_{ba})(D^{\dagger}_{1a\alpha}D^{*}_{2b\lambda\tau}-D_{1b\alpha}
D^{\dagger*}_{2a\lambda\tau})\nonumber\\
&&+\frac{2\lambda^{\prime\prime} g_{V}}{\sqrt{3}}[3\epsilon^{\mu\lambda\nu\tau}v_\lambda(D^{\alpha\dagger}_{1a}
D^{*}_{2b\alpha\tau}+D^{\alpha}_{1b}D^{*\dagger}_{2a\alpha\tau})\partial_\mu \mathbb{V}_{ba\nu}\nonumber\\
&&+2\epsilon^{\lambda\alpha\rho\nu}v_\rho(D^{\dagger}_{1a\alpha}D^{*\mu}_{2b\lambda}
+D_{1b\alpha}D^{\dagger\mu*}_{2a\lambda})\nonumber\\
&&\times(\partial_\mu \mathbb{V}_{ba\nu}-\partial_\nu \mathbb{V}_{ba\mu})].
\end{eqnarray}

In Refs. \cite{Meng:2017fwb,Meng:2017udf}, the effective Lagrangians for $S$-wave doubly charmed baryons with the light mesons are constructed as
\begin{eqnarray}
\label{eq:lagxi}
\mathcal{L}_{\Xi_{cc}}&=& g_{\sigma}\bar{\Xi}_{cc}\sigma \Xi_{cc}+g_{\pi}\bar{\Xi}_{cc}i\gamma_{5}\mathbb{P}\Xi_{cc}\nonumber\\
&&+h_{v}\bar{\Xi}_{cc}\gamma_{\mu}\Xi_{cc}\mathbb{V}^{\mu}
+\frac{f_v}{2M_{\Xi_{cc}}}\bar{\Xi}_{cc}\sigma_{\mu\nu}\partial^{\mu}
\mathbb{V}^{\nu}\Xi_{cc}.
\end{eqnarray}

In addition, the normalization relations for axial-vector meson $D_{1}$, tensor meson $D_{2}^{*}$, and the $S$-wave baryon $\Xi_{cc}$ satisfy
\begin{eqnarray}
\langle 0|D_{1}^{\mu}|Q\bar{q}(1^+)\rangle&=&\epsilon^\mu\sqrt{M_{D_{1}}},\\
\langle 0|D_{2}^{*\mu\nu}|Q\bar{q}(2^+)\rangle&=&\zeta^{\mu\nu}\sqrt{M_{D_{2}^*}},\\
\langle 0|\Xi_{cc}|QQq(1/2^+)\rangle&=&\sqrt{2M_{\Xi_{cc}}}
({\chi,\frac{\bm{\sigma}\cdot\bm{p}}{2M_{\Xi_{cc}}}\chi)^T}.
\end{eqnarray}

So far, the coupling constants in Eqs. (\ref{eq:lagp}) and (\ref{eq:lagxi}) cannot be fixed based on the experimental data. In this work, we will estimate all of the coupling constants in a quark model, where the weak interaction between heavy quarks and light quarks is ignored. For the light quark interactions, they can be extracted from nucleon-nucleon interaction,
\begin{eqnarray}
\mathcal L_{N}&=&g_{\sigma NN}\bar N \sigma N+\sqrt{2}g_{\pi NN}\bar N i\gamma_{5}\pi N\nonumber\\
&&+\sqrt{2}g_{\rho NN}\bar N \gamma_{\mu}\rho^{\mu}N+\frac{f_{\rho NN}}{\sqrt{2}M_N}\bar N \sigma_{\mu\nu} \partial^{\mu}\rho^{\nu}N.
\end{eqnarray}
In Table \ref{Parameters}, we collect the values for all of the coupling constants. The derivations are presented in the Appendix~\ref{app01}. In particular, we need to emphasize that in the quark model \cite{Riska:2000gd}, the coupling $k$ for the $D_1 D_2^* \pi$ vertex is the same as the $D^{*}D\pi$ coupling $g = 0.59\pm0.07\pm0.01$, and the latter is extracted the decay width of the $D^{*+}\rightarrow D^{0}\pi^{+}$ process \cite{Isola:2003fh}. We also adopt $k=g$ in the following calculations.

\renewcommand\tabcolsep{0.3cm}
\renewcommand{\arraystretch}{1.8}
\begin{table*}[!htbp]
\caption{A summary of hadron masses from PDG \cite{Tanabashi:2018oca} and coupling constants adopted in our calculations. Here, masses are taken as the average values, for example, $m_{D_1}=(m_{D_1^{+}}+m_{D_1^{0}})/2$. Hadron masses are given in units of MeV.}\label{Parameters}
\begin{tabular}{llllll}\toprule[1.5pt]\midrule[1.0pt]
            &  $\Xi_{cc}$  &  $T$ doublt   &  $N$  \cite{Machleidt:1987hj,Machleidt:2000ge,Cao:2010km} &$m_\sigma$=600 &$m_{\Xi_{cc}}$=3621.40\\\cline{1-4}
 $\sigma$ exchange &$g_{\sigma}=\frac{1}{3}g_{\sigma NN}$  & $g_{\sigma}^{\prime\prime}=\frac{1}{3}g_{\sigma NN}$&$\frac{g_{\sigma NN}^2}{4\pi}=5.69$ &$m_\pi$=137.24  &$m_{D_1}$=2422.00\\
 $\pi/\eta$ exchange &$g_{\pi}=-\frac{\sqrt{2}M_{\Xi_{cc}}}{5M_N}g_{\pi NN}$ &$\frac{k}{f_\pi}=\frac{3}{5\sqrt{2} M_N}g_{\pi NN}$&$\frac{g_{\pi NN}^2}{4\pi}=13.60$ &$m_\eta$=547.28   &$m_{D_2^*}$=2463.05\\
 $\rho/\omega$ exchange & $h_{v}=\sqrt{2}g_{\rho NN}$  &$\beta^{\prime\prime} g_{V}=-2g_{\rho NN}$ &$\frac{g_{\rho NN}^2}{4\pi}=0.84$  &$m_\rho$=775.49   &$m_N$=938.27\\
 & $h_{v}+f_{v}=-\frac{\sqrt{2}M_{\Xi_{cc}}}{5M_N}(g_{\rho NN}+f_{\rho NN})$ & $\lambda^{\prime\prime} g_{V}=\frac{3}{10M_N}(g_{\rho NN}+f_{\rho NN})$&$\frac{f_{\rho NN}}{g_{\rho NN}}=6.10$ &$m_\omega$=782.65\\
\bottomrule[1.5pt]\midrule[1.0pt]
\end{tabular}
\end{table*}

\subsection{Potentials}

With the help of the Breit approximation, the OBE effective potentials $\mathcal{V}^{h_1h_2\to h_3h_4}(\bm{q})$ in momentum space can be related to the $t$-channel scattering amplitude for process $h_1h_2\to h_3h_4$,
\begin{eqnarray}\label{breit}
\mathcal{V}_E^{h_1h_2\to h_3h_4}(\bm{q}) &=&
          -\frac{\mathcal{M}(h_1h_2\to h_3h_4)}
          {\sqrt{\prod_i2M_i\prod_f2M_f}},
\end{eqnarray}
where $\mathcal{M}(h_1h_2\to h_3h_4)$ stands for the scattering amplitude, $M_i$ and $M_f$ are the masses of the initial states and final states, respectively. Once Fourier transformation is performed, the effective potential $\mathcal{V}(\bm{r})$ in the coordinate space can be deduced, i.e.,
\begin{eqnarray}
\mathcal{V}_E(\bm{r}) =
          \int\frac{d^3\bm{q}}{(2\pi)^3}e^{i\bm{q}\cdot\bm{r}}
          \mathcal{V}_E^{h_1h_2\to h_3h_4}(\bm{q})\mathcal{F}^2(q^2,m_E^2).\label{vr}
\end{eqnarray}
Here, we introduce a form factor $\mathcal{F}(q^2,m_E^2) = (\Lambda^2-m_E^2)/(\Lambda^2-q^2)$ in every interactive vertex. It is used to reflect the finite size effect of the discussed hadrons and compensate the off-shell effects of the exchanged mesons. $\Lambda$, $m_E$, and $q$ are the cutoff, the mass and the four momentum of the exchanged meson, respectively. For the only phenomenological parameter, cutoff $\Lambda$, it is tuned from 0.8 to 5.0 GeV in our following numerical calculations. According to the experience of the deuteron \cite{Tornqvist:1993ng,Tornqvist:1993vu}, the value of the cutoff is taken around 1.0 GeV. A loosely bound state with cutoff around 1.0 GeV can be a possible hadronic molecular candidate.

The OBE effective potentials in the coordinate space for all of the investigated processes are given by
\begin{eqnarray}
\mathcal{V}_1&=&
-g_{\sigma}g_{\sigma}^{\prime\prime}Y(\Lambda,m_\sigma,r)\mathcal{A}_1\nonumber\\
&&-\frac{5g_\pi k}{36f_\pi M_{\Xi_{cc}}}\left[\mathcal{G}(I)Z(\Lambda,m_\pi,r)+\frac{ Z(\Lambda,m_\eta,r)}{6}\right]
\mathcal{A}_3\nonumber\\
&&-\frac{5g_\pi k}{36f_\pi M_{\Xi_{cc}}}\left[\mathcal{G}(I)T(\Lambda,m_\pi,r)+\frac{ T(\Lambda,m_\eta,r)}{6}\right]
\mathcal{A}_2\nonumber\\
&&+\frac{1}{\sqrt{2}}h_v \beta^{\prime\prime} g_{V}\left[\mathcal{G}(I)Y(\Lambda,m_\rho,r)+\frac{Y(\Lambda,m_\omega,r)}{2}\right]
\mathcal{A}_1\nonumber\\
&&-\frac{5h_v \lambda^{\prime\prime} g_V}{3\sqrt{2}M_{\Xi_{cc}}}\left[\mathcal{G}(I)Q(\Lambda,m_\rho,r)
+\frac{Q(\Lambda,m_\omega,r)}{2}\right]\mathcal{A}_4\nonumber\\
&&\!\!\!\!\!\!\!\!\!\!\!\!-\frac{5(h_v+f_v) \lambda^{\prime\prime} g_V}{9\sqrt{2}M_{\Xi_{cc}}}\left[\mathcal{G}(I)Z(\Lambda,m_\rho,r)
+\frac{Z(\Lambda,m_\omega,r)}{2}\right]\mathcal{A}_3\nonumber\\
&&\!\!\!\!\!\!\!\!\!\!\!\!+\frac{5(h_v+f_v) \lambda^{\prime\prime} g_V}{18\sqrt{2}M_{\Xi_{cc}}}
\left[\mathcal{G}(I)T(\Lambda,m_\rho,r)+\frac{ T(\Lambda,m_\omega,r)}{2}\right]\mathcal{A}_2,\nonumber\\\\
%%%%%%%%
\mathcal{V}_2&=&-g_{\sigma}g_{\sigma}^{\prime\prime}
Y(\Lambda,m_\sigma,r)\mathcal{A}_5\nonumber\\
&&-\frac{g_\pi k}{6f_\pi M_{\Xi_{cc}}}\left[\mathcal{G}(I)Z(\Lambda,m_\pi,r)+\frac{ Z(\Lambda,m_\eta,r)}{6}\right]
\mathcal{A}_7\nonumber\\
&&-\frac{g_\pi k}{6f_\pi M_{\Xi_{cc}}}\left[\mathcal{G}(I)T(\Lambda,m_\pi,r)+\frac{ T(\Lambda,m_\eta,r)}{6}\right]
\mathcal{A}_6\nonumber\\
&&+\frac{1}{\sqrt{2}}h_v \beta^{\prime\prime} g_{V}\left[\mathcal{G}(I)Y(\Lambda,m_\rho,r)+\frac{Y(\Lambda,m_\omega,r)}{2}\right]
\mathcal{A}_5\nonumber\\
&&-\frac{\sqrt{2}h_v \lambda^{\prime\prime} g_V}{M_{\Xi_{cc}}}\left[\mathcal{G}(I)Q(\Lambda,m_\rho,r)+\frac{ Q(\Lambda,m_\omega,r)}{2}\right]
\mathcal{A}_8\nonumber\\
&&\!\!\!\!\!\!\!\!-\frac{\sqrt{2}(h_v+f_v) \lambda^{\prime\prime} g_V}{3M_{\Xi_{cc}}}\left[\mathcal{G}(I)Z(\Lambda,m_\rho,r)+\frac{ Z(\Lambda,m_\omega,r)}{2}\right]
\mathcal{A}_7\nonumber\\
&&\!\!\!\!\!\!\!\!+\frac{(h_v+f_v)\lambda^{\prime\prime} g_V}{3\sqrt{2}M_{\Xi_{cc}}}
\left[\mathcal{G}(I)T(\Lambda,m_\rho,r)+\frac{ T(\Lambda,m_\omega,r)}{2}\right]\mathcal{A}_6,\nonumber\\\\
%%%%%%%
\mathcal{V}_3&=&
-\frac{g_\pi k\mathcal{A}_{10}}{3\sqrt{24}f_\pi M_{\Xi_{cc}}}\left[\mathcal{G}(I)Z(\Lambda_0,m_{\pi0},r)+\frac{ Z(\Lambda_0,m_{\eta0},r)}{6}\right]
\nonumber\\
&&-\frac{g_\pi k\mathcal{A}_9}{3\sqrt{24}f_\pi M_{\Xi_{cc}}}
\left[\mathcal{G}(I)T(\Lambda_0,m_{\pi0},r)+\frac{T(\Lambda_0,m_{\eta0},r)}{6}\right]
\nonumber\\
&&-\frac{h_v \lambda^{\prime\prime} g_V}{\sqrt{3}M_{\Xi_{cc}}}\mathcal{A}_{11}
\left[\mathcal{G}(I)Q(\Lambda_0,m_{\rho0},r)+\frac{Q(\Lambda_0,m_{\omega0},r)}{2}\right]
\nonumber\\
&&\!\!\!\!\!\!\!\!-\frac{(h_v+f_v) \lambda^{\prime\prime}g_V}{3\sqrt{3}M_{\Xi_{cc}}}
\left[\mathcal{G}(I)Z(\Lambda_0,m_{\rho0},r)+\frac{Z(\Lambda_0,m_{\omega0},r)}{2}\right]
\mathcal{A}_{10}\nonumber\\
&&\!\!\!\!\!\!\!\!-\frac{(h_v+f_v) \lambda^{\prime\prime}g_V}{6\sqrt{3}M_{\Xi_{cc}}}\left[\mathcal{G}(I)T(\Lambda_0,m_{\rho0},r)
+\frac{T(\Lambda_0,m_{\omega0},r)}{2}\right]\mathcal{A}_9.\nonumber\\
\end{eqnarray}
Here, $\mathcal{V}_1$, $\mathcal{V}_2$, and $\mathcal{V}_3$ correspond to the OBE effective potentials for processes $\Xi_{cc}D_1\to\Xi_{cc}D_1$, $\Xi_{cc}D_2^*\to\Xi_{cc}D_2^*$, and $\Xi_{cc}D_1\to\Xi_{cc}D_2^*$, respectively. Functions $Y(\Lambda,m,r)$, $T(\Lambda,m,r)$, $Z(\Lambda,m,r)$, and $Q(\Lambda,m,r)$ are defined as
\begin{eqnarray}
Y(\Lambda,m,r)&=&\frac{1}{4\pi r}(e^{-mr}-e^{-\Lambda r})-\frac{\Lambda^2-m^2}{8 \pi\Lambda}e^{-\Lambda r},\\
T(\Lambda,m,r)&=&r\frac{\partial}{\partial r}\frac{1}{r}\frac{\partial}{\partial r}Y(\Lambda,m,r),\\
Z(\Lambda,m,r)&=&\nabla^2Y(\Lambda,m,r),\\
Q(\Lambda,m,r)&=&\frac{1}{r}\frac{\partial}{\partial r}Y(\Lambda,m,r).
\end{eqnarray}
Variables in the above effective potentials denote
\begin{eqnarray}
\Lambda_0^2=\Lambda^2-q_0^2,\quad
m_{E0}^2=m_{E}^2-q_0^2,\quad
q_0=\frac{m_{D_2^{*}}^2-m_{D_1}^2}{2(m_{\Xi_{cc}}+m_{D_2^{*}})}.\nonumber
\end{eqnarray}
$I$ stands for the isospin for $\Xi_{cc}D_1/\Xi_{cc}D_2^*$ systems. $\mathcal{G}(I)$ is the isospin factor: $\mathcal{G}(I=0)=3/2$ and $\mathcal{G}(I=1)=-1/2$.

In the above OBE effective potentials, we also introduce several spin-spin, spin-orbit, and tensor force operators,
\begin{eqnarray*}
\label{111}
\mathcal{A}_{1}&=&{\bm\epsilon^{\dagger}_4}\cdot{\bm\epsilon_2}\chi^{\dagger}_3\chi_1,
\quad\quad\quad\quad\quad
\mathcal{A}_{2}=\chi^{\dagger}_3S({\bm\sigma_1},i{\bm\epsilon^{\dagger}_4}
\times{\bm\epsilon_2},\hat{\bm{r}})\chi_1,\\
\mathcal{A}_{3}&=&\chi^{\dagger}_3 {\bm\sigma_1}\cdot(i{\bm\epsilon^{\dagger}_4}\times {\bm\epsilon_2})\chi_1,\quad\quad
\mathcal{A}_{4}=(i{\bm\epsilon^{\dagger}_4}\times {\bm\epsilon_2})\cdot{\bm L}\chi^{\dagger}_3\chi_1,\\
\mathcal{A}_{5}&=&\sum_{m,n,a,b}C^{2,m+n}_{1m,1n}C^{2,a+b}_{1a,1b}
({\bm\epsilon^{\dagger}_{4m}}\cdot{\bm\epsilon_{2a}})({\bm\epsilon^{\dagger}_{4n}}\cdot {\bm\epsilon_{2b}})\chi^{\dagger}_3\chi_1,\\
\mathcal{A}_{6}&=&\sum_{m,n,a,b}C^{2,m+n}_{1m,1n}C^{2,a+b}_{1a,1b}
({\bm\epsilon^{\dagger}_{4m}}\cdot{\bm\epsilon_{2a}})\chi^{\dagger}_3S({\bm\sigma_1},i{\bm \epsilon^{\dagger}_{4n}}\times{\bm\epsilon_{2b}},\hat{\bm{r}})\chi_1,\\
\mathcal{A}_{7}&=&\sum_{m,n,a,b}C^{2,m+n}_{1m,1n}C^{2,a+b}_{1a,1b}
({\bm\epsilon^{\dagger}_{4m}}\cdot{\bm\epsilon_{2a}})\chi^{\dagger}_3{\bm\sigma_1}\cdot
(i{\bm\epsilon^{\dagger}_{4n}}\times {\bm\epsilon_{2b}})\chi_1,\\
\mathcal{A}_{8}&=&\sum_{m,n,a,b}C^{2,m+n}_{1m,1n}C^{2,a+b}_{1a,1b}
({\bm\epsilon^{\dagger}_{4m}}\cdot{\bm\epsilon_{2a}})(i{\bm\epsilon^{\dagger}_{4n}}\times {\bm\epsilon_{2b}})\cdot{\bm L}\chi^{\dagger}_3\chi_1,\\
\mathcal{A}_{9}&=&\sum_{m,n}C^{2,m+n}_{1m,1n}({\bm\epsilon^{\dagger}_{4m}}
\cdot{\bm\epsilon_{2}})\chi^{\dagger}_3S({\bm\sigma_1},
{\bm\epsilon^{\dagger}_{4n}},\hat{\bm{r}})\chi_1,\\
\mathcal{A}_{10}&=&\sum_{m,n}C^{2,m+n}_{1m,1n}({\bm\epsilon^{\dagger}_{4m}}
\cdot{\bm\epsilon_{2}})\chi^{\dagger}_3({\bm\sigma_1}\cdot
{\bm\epsilon^{\dagger}_{4n}})\chi_1,\\
\mathcal{A}_{11}&=&\sum_{m,n}C^{2,m+n}_{1m,1n}({\bm\epsilon^{\dagger}_{4m}}
\cdot{\bm\epsilon_{2}})({\bm\epsilon^{\dagger}_{4n}}\cdot{\bm
L})\chi^{\dagger}_3\chi_1.
\end{eqnarray*}
Here, $S({\bm x},{\bm y},\hat{\bm r})$ is the tensor force operator,
\begin{equation}
S({\bm x},{\bm y},\hat{\bm r})= 3(\hat{\bm r} \cdot {\bm x})(\hat{\bm r} \cdot {\bm y})-{\bm x} \cdot {\bm y},
\end{equation}
with $\hat{\bm r}={\bm r}/|{\bm r}|$. In Table \ref{operator1}, we collect the corresponding numerical matrices for these operators.
\renewcommand\tabcolsep{0.70cm}
\renewcommand{\arraystretch}{1.8}
\begin{table*}[!htbp]
\caption{Matrix elements for the angular momentum operators.}\label{operator1}
{\begin{tabular}{cccc}\toprule[1.5pt]\midrule[1.5pt]
$\left\langle \Xi_{cc}D_{1}|\mathcal{A}_{1}|\Xi_{cc}D_{1}\right\rangle_{J=1/2}$ &$\left\langle \Xi_{cc}D_{1}|\mathcal{A}_{2}|\Xi_{cc}D_{1}\right\rangle_{J=1/2}$&$\left\langle \Xi_{cc}D_{1}|\mathcal{A}_{3}|\Xi_{cc}D_{1}\right\rangle_{J=1/2}$&$\left\langle \Xi_{cc}D_{1}|\mathcal{A}_{4}|\Xi_{cc}D_{1}\right\rangle_{J=1/2}$ \\
$\left(\begin{array}{cc} 1 &0 \\ 0 &1\end{array}\right)$&$\left(\begin{array}{cc} 0 &\sqrt{2} \\ \sqrt{2} &2\end{array}\right)$ & $\left(\begin{array}{cc} 2&0 \\ 0 &-1 \end{array}\right)$ &$\left(\begin{array}{cc} 0 &0 \\ 0 &3\end{array}\right)$\\
$\left\langle \Xi_{cc}D_{2}^{*}|\mathcal{A}_{5}|\Xi_{cc}D_{2}^{*}\right\rangle_{J=1/2}$&$\left\langle \Xi_{cc}D_{2}^{*}|\mathcal{A}_{6}|\Xi_{cc}D_{2}^{*}\right\rangle_{J=1/2}$&$\left\langle \Xi_{cc}D_{2}^{*}|\mathcal{A}_{7}|\Xi_{cc}D_{2}^{*}\right\rangle_{J=1/2}$&$\left\langle \Xi_{cc}D_{2}^{*}|\mathcal{A}_{8}|\Xi_{cc}D_{2}^{*}\right\rangle_{J=1/2}$\\
$\left(\begin{array}{cc} 1&0 \\ 0 &1 \end{array}\right)$ &$\left(\begin{array}{cc} -\frac{3}{5}&\frac{3\sqrt{6}}{10}\\ \frac{3\sqrt{6}}{10} &\frac{8}{5}\end{array}\right)$&$\left(\begin{array}{cc} \frac{3}{2} &0 \\ 0  &-1\end{array}\right)$ & $\left(\begin{array}{cc} \frac{27}{10} &-\frac{\sqrt{6}}{10} \\-\frac{\sqrt{6}}{10}&\frac{14}{5}\end{array}\right)$\\
$\left\langle \Xi_{cc}D_{2}^{*}|\mathcal{A}_{9}|\Xi_{cc}D_{1}\right\rangle_{J=1/2}$&$\left\langle \Xi_{cc}D_{2}^{*}|\mathcal{A}_{10}|\Xi_{cc}D_{1}\right\rangle_{J=1/2}$&$\left\langle \Xi_{cc}D_{2}^{*}|\mathcal{A}_{11}|\Xi_{cc}D_{1}\right\rangle_{J=1/2}$\\
$\left(\begin{array}{cc} \frac{1}{\sqrt{5}}&\frac{2\sqrt{6}}{\sqrt{5}} \\ \sqrt{\frac{2}{5}} &-\sqrt{\frac{3}{5}} \end{array}\right)$ &$\left(\begin{array}{cc} 0 &0 \\ \sqrt{\frac{5}{2}}  &0\end{array}\right)$&$\left(\begin{array}{cc} 0 &0 \\ -\frac{3}{\sqrt{10}}  &-\sqrt{\frac{3}{5}}\end{array}\right)$\\
$\left\langle \Xi_{cc}D_{1}|\mathcal{A}_{1}|\Xi_{cc}D_{1}\right\rangle_{J=3/2}$ &$\left\langle \Xi_{cc}D_{1}|\mathcal{A}_{2}|\Xi_{cc}D_{1}\right\rangle_{J=3/2}$&$\left\langle \Xi_{cc}D_{1}|\mathcal{A}_{3}|\Xi_{cc}D_{1}\right\rangle_{J=3/2}$&$\left\langle \Xi_{cc}D_{1}|\mathcal{A}_{4}|\Xi_{cc}D_{1}\right\rangle_{J=3/2}$\\
$\left(\begin{array}{ccc} 1 &0&0 \\ 0 &1&0\\ 0&0&1\end{array}\right)$&$\left(\begin{array}{ccc} 0 &-1&-2 \\ -1 &0&1\\ -2&1&0\end{array}\right)$&$\left(\begin{array}{ccc} -1 &0&0 \\0 &2&0\\ 0&0&-1\end{array}\right)$&$\left(\begin{array}{ccc} 0 &0&0 \\0 &2&-1\\ 0&-1&2\end{array}\right)$\\
$\left\langle \Xi_{cc}D_{2}^{*}|\mathcal{A}_{5}|\Xi_{cc}D_{2}^{*}\right\rangle_{J=3/2}$&$\left\langle \Xi_{cc}D_{2}^{*}|\mathcal{A}_{6}|\Xi_{cc}D_{2}^{*}\right\rangle_{J=3/2}$&$\left\langle \Xi_{cc}D_{2}^{*}|\mathcal{A}_{7}|\Xi_{cc}D_{2}^{*}\right\rangle_{J=3/2}$&$\left\langle \Xi_{cc}D_{2}^{*}|\mathcal{A}_{8}|\Xi_{cc}D_{2}^{*}\right\rangle_{J=3/2}$\\
$\left(\begin{array}{ccc} 1 &0&0 \\ 0 &1&0\\ 0&0&1\end{array}\right)$&$\left(\begin{array}{ccc} 0 &\frac{3}{5}&\frac{3\sqrt{21}}{10} \\ \frac{3}{5} &0&\frac{3\sqrt{21}}{14} \\ \frac{3\sqrt{21}}{10} &\frac{3\sqrt{21}}{14}&\frac{4}{7}\end{array}\right)$&$\left(\begin{array}{ccc} \frac{3}{2} &0&0 \\ 0 &\frac{3}{2}&0\\ 0&0&-1\end{array}\right)$&$\left(\begin{array}{ccc} 0&0&0 \\ 0 &\frac{9}{5}&-\frac{\sqrt{21}}{10}\\ 0&-\frac{\sqrt{21}}{10}&\frac{11}{5}\end{array}\right)$\\
$\left\langle \Xi_{cc}D_{2}^{*}|\mathcal{A}_{9}|\Xi_{cc}D_{1}\right\rangle_{J=3/2}$&$\left\langle \Xi_{cc}D_{2}^{*}|\mathcal{A}_{10}|\Xi_{cc}D_{1}\right\rangle_{J=3/2}$&$\left\langle \Xi_{cc}D_{2}^{*}|\mathcal{A}_{11}|\Xi_{cc}D_{1}\right\rangle_{J=3/2}$\\
$\left(\begin{array}{ccc} 0 &-\sqrt{\frac{2}{5}}  &-\sqrt{\frac{21}{10}} \\ -\frac{1}{\sqrt{10}} &\frac{1}{\sqrt{10}}    & -\frac{2\sqrt{6}}{\sqrt{35}} \\ -\sqrt{\frac{2}{5}} &0   &-\sqrt{\frac{15}{14}}  \end{array}\right)$&$\left(\begin{array}{ccc} \sqrt{\frac{5}{2}}   &0 &0\\ 0 & 0 & 0\\ 0 & \sqrt{\frac{5}{2}}  & 0 \end{array}\right)$&$\left(\begin{array}{ccc} 0 &0 &0\\   0   & -\sqrt{\frac{5}{2}}  & 0 \\ 0   &-\sqrt{\frac{2}{5}}    &-\sqrt{\frac{21}{10}} \end{array}\right)$ \\
\bottomrule[1.5pt]\midrule[1.5pt]
\end{tabular}}
\end{table*}

\section{Numerical results}\label{sec4}

After we prepared the OBE effective potentials in Sec. \ref{sec3}, here we test if the $\Xi_{cc}D_{1}$ and $\Xi_{cc}D_{2}^{*}$ systems can be possible triple-charm pentaquark molecules by numerically solving the Schr\"{o}dinger equation,
\begin{eqnarray}
-\frac{1}{2\mu}\Big(\nabla^2-\frac{l(l+1)}{r^2}\Big)\psi(r)+V(r)\psi(r)=E\psi(r),
\end{eqnarray}
where $\nabla^2=\frac{1}{r^2}\frac{\partial}{\partial r}r^2\frac{\partial}{\partial r}$, and $\mu=\frac{m_1m_2}{m_1+m_2}$ being the reduced mass for the discussed systems.

Before we present our numerical results, let us briefly emphasize the cutoff parameter $\Lambda$ introduced in the form factor. Its value is related to the typical hadronic scale or to the intrinsic size of hadrons; the reasonable values should be around 1.0 GeV ~\cite{Tornqvist:1993ng,Tornqvist:1993vu}. Although the bound state properties depend on the choice of $\Lambda$, we can predict possible loosely bound molecular states at the qualitative level.

\subsection{Single $\Xi_{cc}D_{1}$ and $\Xi_{cc}D_{2}^{*}$ systems}\label{sub1}

\begin{figure}[!htbp]
\centering
\begin{tabular}{c}
\includegraphics[width=0.43\textwidth]{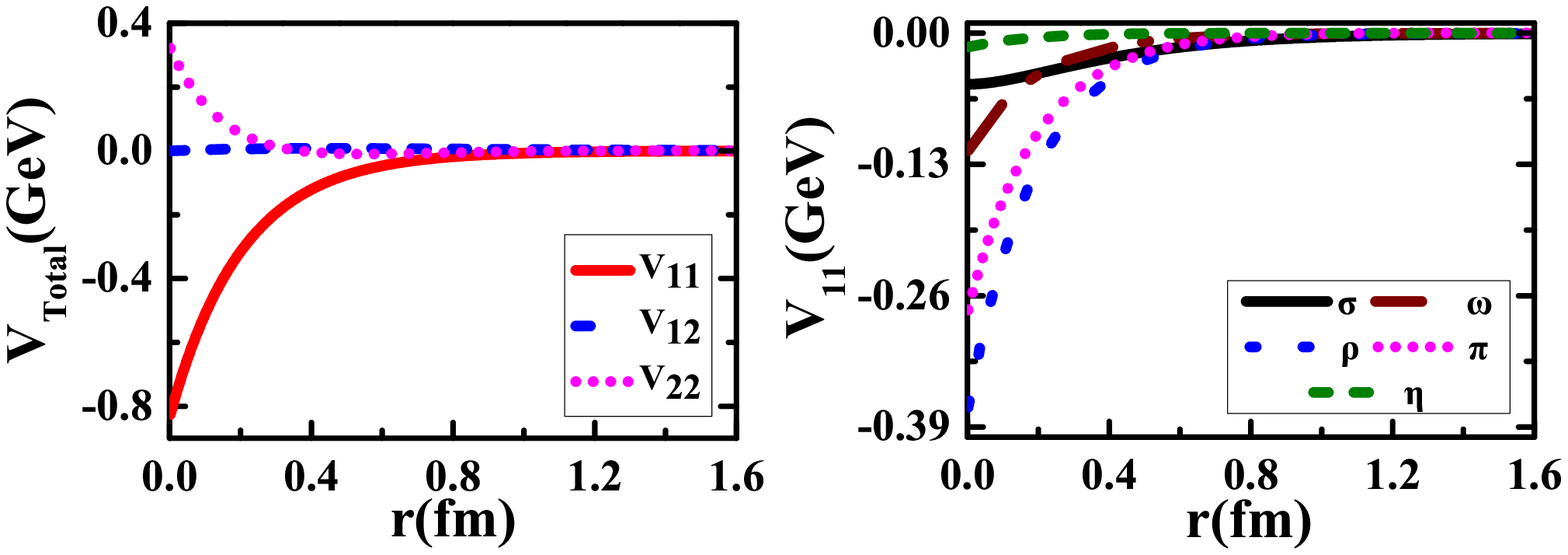}\\
\includegraphics[width=0.43\textwidth]{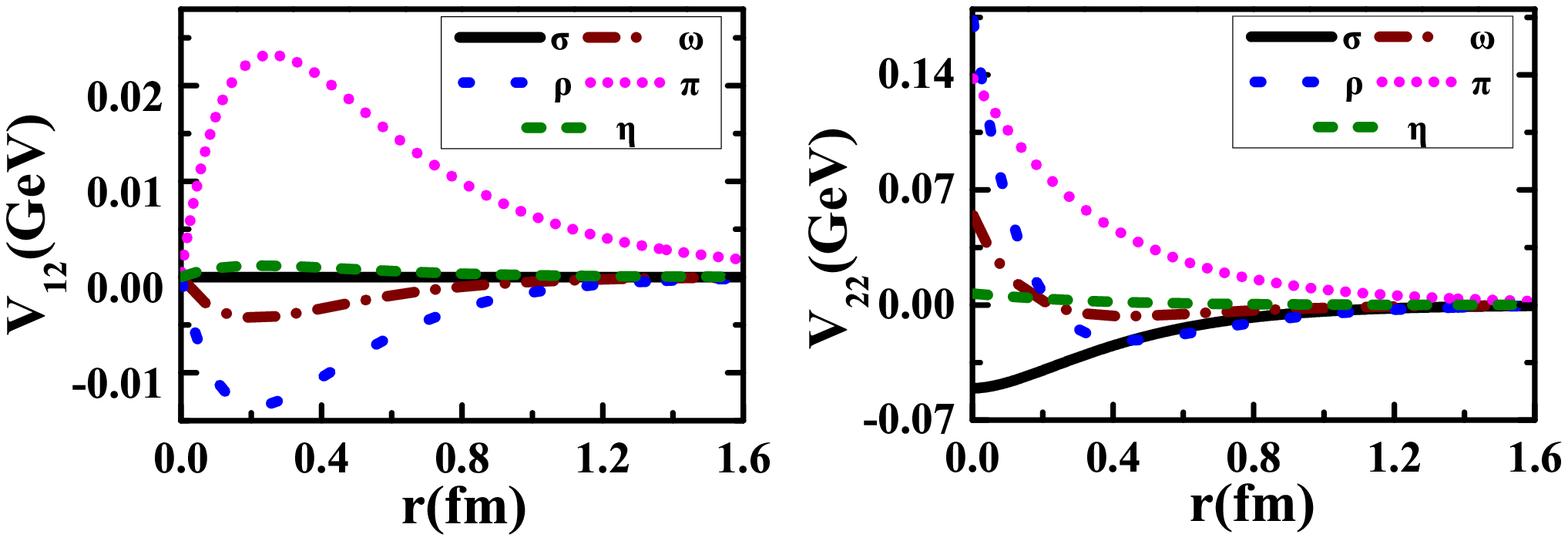}
\end{tabular}
\caption{The effective potential for the $\Xi_{cc}D_{1}$ system with $I(J^P)=0(1/2^+)$, and cutoff $\Lambda$ is fixed as $1.00$ GeV. Here, $V_{11}({\bf{r}})=\langle{}^2\mathbb{S}_{\frac{1}{2}}|\mathcal{V}^{\Xi_{cc}D_{1}
\rightarrow\Xi_{cc}D_{1}}({\bf{r}})|{}^2\mathbb{S}_{\frac{1}{2}}\rangle$, $V_{12}({\bf{r}})=\langle{}^4\mathbb{D}_{\frac{1}{2}}|\mathcal{V}^{\Xi_{cc}D_{1}
\rightarrow\Xi_{cc}D_{1}}({\bf{r}})|{}^2\mathbb{S}_{\frac{1}{2}}\rangle$, and $V_{22}({\bf{r}})=\langle{}^4\mathbb{D}_{\frac{1}{2}}|\mathcal{V}^{\Xi_{cc}D_{1}
\rightarrow\Xi_{cc}D_{1}}({\bf{r}})|{}^4\mathbb{D}_{\frac{1}{2}}\rangle$. }\label{potentialshape}
\end{figure}

For the $\Xi_{cc}D_{1}$ and $\Xi_{cc}D_{2}^{*}$ systems, $\pi/\eta/\sigma/\rho/\omega$ exchanges are allowed. According to the masses of the exchanged mesons, $\pi$ exchange, $\sigma/\eta$ exchanges, and $\rho/\omega$ exchanges contribute to the long, intermediate, and short range forces, respectively. In Fig. \ref{potentialshape}, we present the $r$ dependence of the OBE effective potentials for the $\Xi_{cc}D_{1}$ system with $I=0,J=1/2$, and the cutoff $\Lambda$ is taken as $1.00$ GeV. Here, we notice that:
\begin{itemize}
  \item $\pi$-exchange interaction acts as a typical long range force. When $r>1.2$ fm, only $\pi$-exchange contribution survives.
  \item the properties for the center force provided by scalar and vector meson exchanges are consistent with our previous conclusions, where the $\sigma$-exchange and $\omega$-exchange interactions are attractive, and the interaction from $\rho$ exchange is about 3 times stronger than that provided by $\omega$ exchange.
\end{itemize}

Here, we shall emphasize again the proposals from our HQSS analysis: the interactions for the $\Xi_{cc}D_1$ systems with $I(J^P)=$ $0(1/2^+,\,3/2^+)$ and the $\Xi_{cc}D_2^*$ systems with $I(J^P)=$ $0(3/2^+,\,5/2^+)$ are strongly attractive, and thus they may be possible molecular state candidates. In comparison to those high spin states, the $\Xi_{cc}D_1$ state with $I(J^P)=0(1/2^+)$ and $\Xi_{cc}D_2^*$ state with $I(J^P)=0(3/2^+)$ may be much more tight.

\renewcommand\tabcolsep{0.08cm}
\renewcommand{\arraystretch}{1.8}
\begin{table}[!htbp]
\caption{Bound state solutions for the $\Xi_{cc}D_{1}$ and $\Xi_{cc}D_{2}^{*}$ systems. Here, the cutoff $\Lambda$, the binding energy $E$ , and the root-mean-square radius $r_{\rm rms}$ are in units of $ \rm {GeV}$, $\rm {MeV}$ and , $\rm {fm}$, respectively. }\label{jg1}
\begin{tabular}{cccccccc}\toprule[1.0pt]\midrule[1.0pt]
States&$\rm \Lambda$ &$\rm E$  &$\rm r_{\rm rms}$ &States&$\rm \Lambda$ &$\rm E$ &$\rm r_{\rm rms}$ \\\midrule[1.0pt]
$[\Xi_{cc}D_{1}]_{J=1/2}^{I=0}$&0.91&$-0.87$ &3.03&$[\Xi_{cc}D_{2}^{*}]_{J=3/2}^{I=0}$&0.91&$-0.40$ &4.33 \\
&0.93&$-3.29$&1.69&&0.94&$-3.67$&1.62 \\
&0.96&$-10.30$&1.07& &0.97&$-10.70$ &1.05\\
$[\Xi_{cc}D_{1}]_{J=1/2}^{I=1}$&2.70&$-0.39$&4.52&$[\Xi_{cc}D_{2}^{*}]_{J=3/2}^{I=1}$&2.50&$-0.37$ &4.58\\
&3.85&$-1.82$&2.23&&3.75&$-2.37$&1.98\\
&5.00&$-3.02$&1.79&&5.00&$-3.94$ &1.59\\
$[\Xi_{cc}D_{1}]_{J=3/2}^{I=0}$&1.09&$-0.36$ &4.77&$[\Xi_{cc}D_{2}^{*}]_{J=5/2}^{I=0}$&1.11&$-0.49$ &4.18\\
&1.23&$-4.75$&1.63&&1.23&$-5.12$&1.60 \\
&1.31&$-12.36$&1.16&&1.34&$-12.11$&1.19\\
$[\Xi_{cc}D_{1}]_{J=3/2}^{I=1}$&1.62&$-0.30$ &5.00&$[\Xi_{cc}D_{2}^{*}]_{J=5/2}^{I=1}$&1.58&$-0.34$ &4.72 \\
&2.09&$-4.66$&1.47&&2.00&$-4.59$ &1.47\\
&2.56&$-11.60$ &1.00&&2.42&$-11.50$ &1.00 \\
\bottomrule[1.0pt]\midrule[1.0pt]
\end{tabular}
\end{table}

\begin{figure}[!htbp]
\centering
\begin{tabular}{c}
\includegraphics[width=0.43\textwidth]{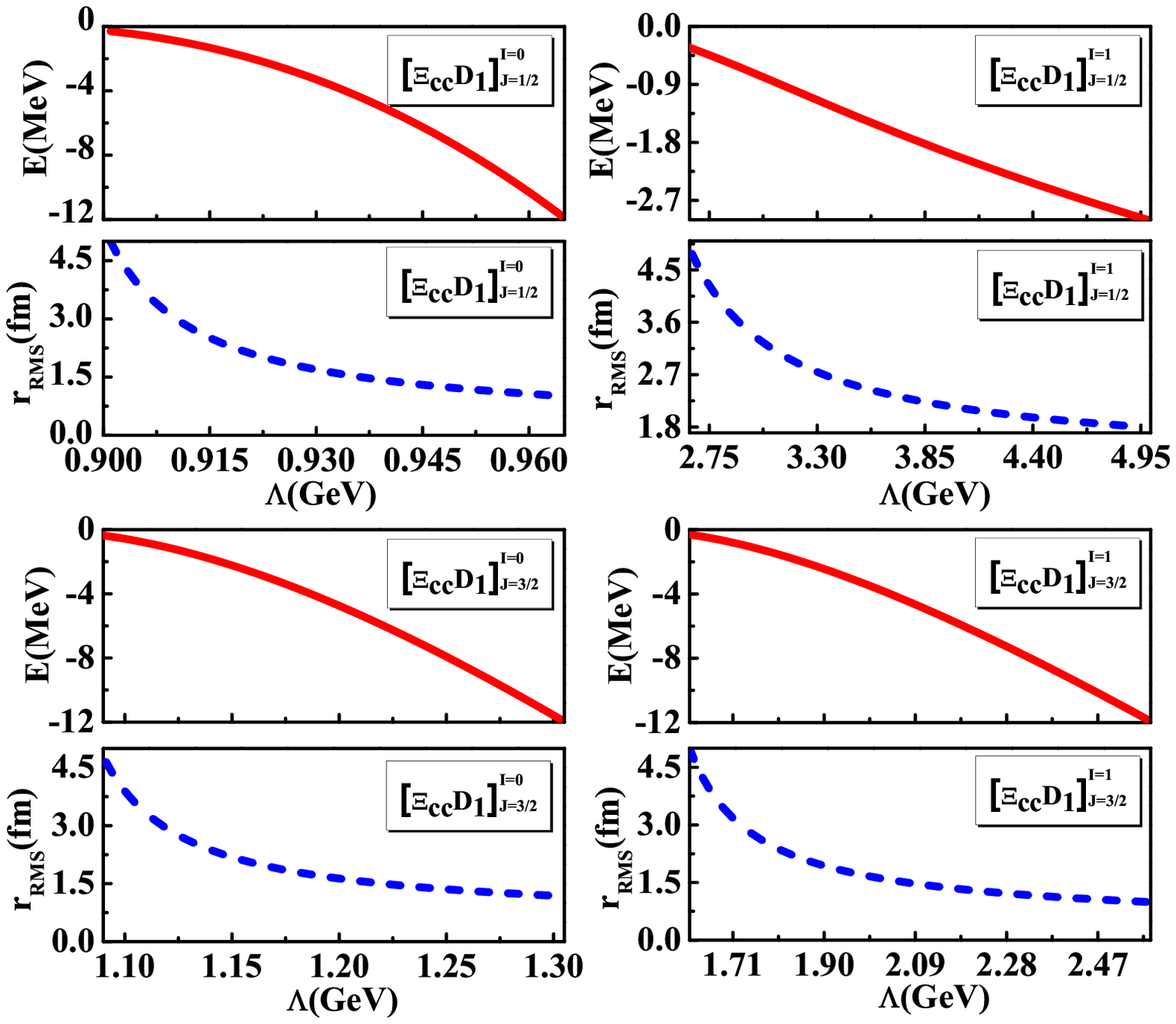}\\
\includegraphics[width=0.43\textwidth]{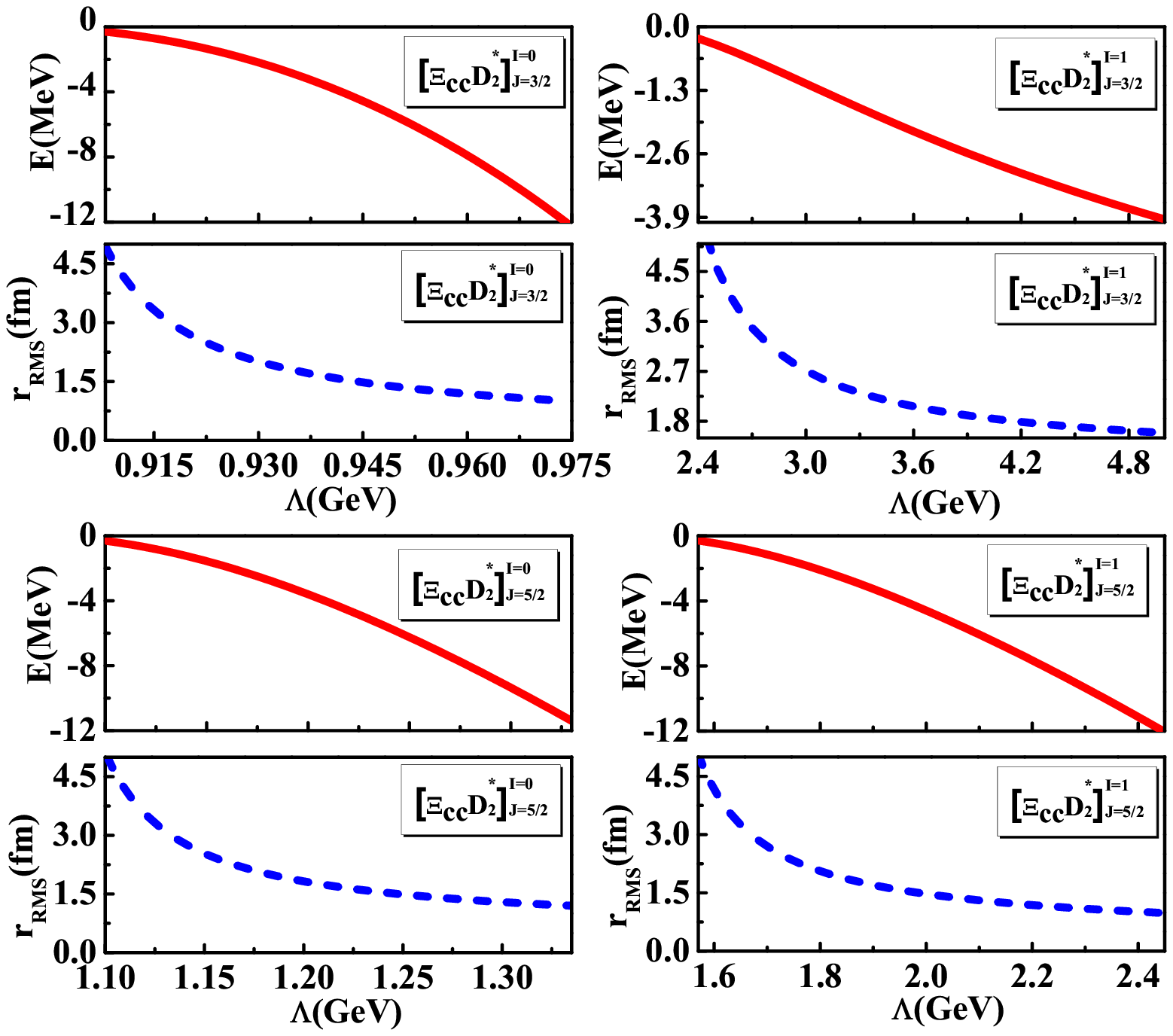}
\end{tabular}
\caption{$\Lambda$ dependence of bound state solutions(the binding energy $E$ and root-mean-square radius $r_{rms}$) for the single $\Xi_{cc} D_{1}$ and $\Xi_{cc} D_{2}^{*}$ systems with all the possible configurations. }\label{charmed}
\end{figure}

In Table \ref{jg1}, we collect the bound properties for the $\Xi_{cc}D_{1}$ and $\Xi_{cc}D_{2}^{*}$ systems with different quantum number configurations, where the $\Lambda$ dependence of the binding energy $E$ and the root-mean-square radius $r_{rms}$ for the $\Xi_{cc} D_{1}$ and $\Xi_{cc} D_{2}^{*}$ systems is presented in Fig.~\ref{charmed}. When we tuned cutoff $\Lambda$ from 0.8 to 5.0 GeV, we can obtain bound solutions for all of the investigated systems, the $\Xi_{cc}D_1$ systems with $I(J^P)=$ $0(1/2^+,\,3/2^+)$, $1(1/2^+,\,3/2^+)$, and the $\Xi_{cc}D_2^*$ systems with $I(J^P)=$ $0(3/2^+,\,5/2^+)$, $1(3/2^+,\,5/2^+)$. If we strictly take the cutoff $\Lambda$ around 1.0 GeV according to the experience of the deuteron \cite{Tornqvist:1993ng,Tornqvist:1993vu}, the $\Xi_{cc}D_1$ systems with $I(J^P)=$ $0(1/2^+,\,3/2^+)$ and the $\Xi_{cc}D_2^*$ systems with $I(J^P)=$ $0(3/2^+,\,5/2^+)$ can be good triple-charm molecular candidates. The above results also confirm the previous HQSS analysis in the quantitative level.

Additionally, when taking the same binding energy for all the bound states, we obtain two relations, $\Lambda\left([\Xi_{cc}D_1]_{J=1/2}^{I=0}\right)
<\Lambda\left([\Xi_{cc}D_1]_{J=3/2}^{I=0}\right)$
and $\Lambda\left([\Xi_{cc}D_2^*]_{J=3/2}^{I=0}\right)
<\Lambda\left([\Xi_{cc}D_2^*]_{J=5/2}^{I=0}\right)$. In general, a bound state with smaller cutoff value corresponds to a more attractive interaction. Thus, these two cutoff relations also prove our previous estimations in a sense,
\begin{eqnarray}
&&V_{\Xi_{cc}D_1}^{I(J^P)=0(1/2^+)}<V_{\Xi_{cc}D_1}^{I(J^P)=0(3/2^+)}<0,\\ &&V_{\Xi_{cc}D_2^*}^{I(J^P)=0(3/2^+)}<V_{\Xi_{cc}D_2^*}^{I(J^P)=0(5/2^+)}<0.
\end{eqnarray}

To summarize, we can predict that there exist four triple-charm molecular pentaquarks, the $\Xi_{cc}D_1$ systems with $I(J^P)=$ $0(1/2^+,\,3/2^+)$ and the $\Xi_{cc}D_2^*$ systems with $I(J^P)=$ $0(3/2^+,\,5/2^+)$. For the isovector $\Xi_{cc}D_1$ and $\Xi_{cc}D_2^*$ systems, if cutoff $\Lambda$ smaller than 2.0 GeV is a reasonable input, the $\Xi_{cc}D_1$ state with $I(J^P)=1(3/2^+)$ and the $\Xi_{cc}D_2^*$ state with $I(J^P)=1(5/2^+)$ may be possible triple-charm molecular candidates. In addition, their allowed strong decay channels include
$\Omega_{ccc}\sigma$, $\Omega_{ccc}\rho$, $\Xi_{cc}D$, $\Xi_{cc}D^{*}$, and $\Omega_{ccc}\pi\pi$.

\subsection{$\Xi_{cc}D_{1}/\Xi_{cc}D_{2}^{*}$ coupled systems}\label{sub2}

Since the mass of the $\Xi_{cc}D_1$ and $\Xi_{cc}D_2^*$ systems are very close, it is very necessary to consider the coupled channel effect. In this subsection, we discuss the $\Xi_{cc}D_{1}/\Xi_{cc}D_{2}^{*}$ states with $I = 0, 1$ and $J = 1/2, 3/2$ by considering the coupled channel effect, where possible coupled channels are listed as follows:
\begin{eqnarray*}
&&1/2^+:\,\Xi_{cc}D_{1}|{}^2\mathbb{S}_{\frac{1}{2}}/{}^4\mathbb{D}_{\frac{1}{2}}\rangle, \quad\Xi_{cc}D_{2}^{*}|{}^4\mathbb{D}_{\frac{1}{2}}/{}^6\mathbb{D}_{\frac{1}{2}}\rangle,\\
&&3/2^+:\,\Xi_{cc}D_{1}|{}^4\mathbb{S}_{\frac{3}{2}}/{}^2\mathbb{D}_{\frac{3}{2}}
/{}^4\mathbb{D}_{\frac{3}{2}}\rangle, \quad\Xi_{cc}D_{2}^{*}|{}^4\mathbb{S}_{\frac{3}{2}}/
{}^4\mathbb{D}_{\frac{3}{2}}/{}^6\mathbb{D}_{\frac{3}{2}}\rangle.
\end{eqnarray*}

The relevant numerical results for the $\Xi_{cc}D_{1}/\Xi_{cc}D_{2}^{*}$ coupled systems with $I(J^P) =0,1(1/2^+,\,3/2^+)$ are given in Table \ref{jg3}. Here, the cutoff $\Lambda$ is also taken in the range from $0.80$ to $5.00$ GeV.

\renewcommand\tabcolsep{0.19cm}
\renewcommand{\arraystretch}{1.8}
\begin{table*}[!htbp]
\caption{Bound state solutions for the $\Xi_{cc}D_{1}/\Xi_{cc}D_{2}^{*}$ states with $I(J^P) =0,1(1/2^+,\,3/2^+)$. Cutoff $\Lambda$, binding energy $E$ , and root-mean-square radius $r_{rms}$ are in units of $ \rm{GeV}$, $\rm {MeV}$, and $\rm {fm}$, respectively. $P(\%)$ denotes the probability for the different channels. Here, we label the probability for the corresponding channel in a bold manner.}\label{jg3}
\begin{tabular}{cccccccccc}\toprule[1.0pt]\midrule[1.0pt]
$(I,J^P)$&$\Lambda$ &$E$  &$ r_{rms}$ &P($\Xi_{cc}D_{1}|{}^2\mathbb{S}_{\frac{1}{2}}\rangle$)&P($\Xi_{cc}D_{1}|{}^2\mathbb{D}_{\frac{1}{2}}\rangle$)
&P($\Xi_{cc}D_{2}^{*}|{}^4\mathbb{D}_{\frac{1}{2}}\rangle$) &P($\Xi_{cc}D_{2}^{*}|{}^6\mathbb{D}_{\frac{1}{2}}\rangle$) & &  \\\midrule[1.0pt]
$(0,\frac{1}{2}^+)$&0.90&$-0.47$ &3.84&\textbf{99.51}&0.43&$o(10^{-3})$&$0.05$ & &\\
&0.93&$-3.76$ &1.60&\textbf{99.44}&0.46&$o(10^{-3})$&0.09& & \\
&0.96&$-10.78$ &1.04&\textbf{99.59}&0.32&$o(10^{-3})$&0.09 & &\\
$(1,\frac{1}{2}^+)$&2.30&$-0.33$ &4.41&\textbf{99.62}&0.33&$o(10^{-3})$&$0.05$ & &\\
&3.15&$-3.65$ &1.63&\textbf{98.46}&1.19&0.03&0.32& & \\
&4.00&$-10.24$ &1.04&\textbf{96.70}&2.30&0.09&0.92 & &\\
$(I,J^P)$&$\Lambda$ &$E$  &$r_{rms}$&P($\Xi_{cc}D_{1}|{}^4\mathbb{S}_{\frac{3}{2}}\rangle$)&P($\Xi_{cc}D_{1}|{}^2\mathbb{D}_{\frac{3}{2}}\rangle$)
&P($\Xi_{cc}D_{1}|{}^4\mathbb{D}_{\frac{3}{2}}\rangle$)&P($\Xi_{cc}D_{2}^{*}|{}^4\mathbb{S}_{\frac{3}{2}}\rangle$)
&P($\Xi_{cc}D_{2}^{*}|{}^4\mathbb{D}_{\frac{3}{2}}\rangle$) &P($\Xi_{cc}D_{2}^{*}|{}^6\mathbb{D}_{\frac{3}{2}}\rangle$) \\
$(0,\frac{3}{2}^+)$&1.00&$-0.53$ &3.64&\textbf{88.65}&0.17&0.81&10.35&$o(10^{-3})$&$o(10^{-3})$\\
&1.01&$-2.44$ &1.68&\textbf{65.67}&0.14&0.67&\textbf{33.50}&$o(10^{-3})$& $o(10^{-3})$ \\
&1.02&$-6.49$ &0.96&\textbf{41.91}&0.08&0.34&\textbf{57.65}&$o(10^{-3})$&0.01\\
$(1,\frac{3}{2}^+)$&1.50&$-0.22$ &4.87&\textbf{99.36}&0.05&0.24&0.29& $o(10^{-3})$&0.01\\
&1.63&$-1.67$ &2.23&\textbf{95.76}&0.07&0.44&3.50& 0.17&0.06\\
&1.76&$-6.95$ &0.99&\textbf{57.95}&0.04&0.29&\textbf{40.91}&0.44& 0.37\\
\bottomrule[1.0pt]\midrule[1.0pt]
\end{tabular}
\end{table*}

Since a state with a higher partial wave is less likely to form a bound state as its repulsive centrifugal force $l(l+1)/2Mr^2$. For the $\Xi_{cc}D_1/\Xi_{cc}D_2^*$ coupled system with $J^P=1/2^+$, there is only one $S$-wave component, the $\Xi_{cc}D_1|{}^2\mathbb{S}_{\frac{1}{2}}\rangle$. Compared to the single case, bound state properties for the $\Xi_{cc}D_1/\Xi_{cc}D_2^*$ coupled system with $J^P=1/2^+$ do not change too much. As presented in Table \ref{jg3}, if we take the same cutoff value in both the single system and coupled system, binding energies for the $\Xi_{cc}D_1/\Xi_{cc}D_2^*$ coupled systems with $I(J^P)=0(1/2^+)$ increase less than 1.0 MeV. For the isovector $\Xi_{cc}D_1/\Xi_{cc}D_2^*$ coupled systems with $J^P=1/2^+$, this increased bind energy reaches several MeV. All in all, the coupled channel effect plays a positive but minor effect in the $\Xi_{cc}D_1/\Xi_{cc}D_2^*$ coupled systems with $J^P=1/2^+$.

For the $\Xi_{cc}D_1/\Xi_{cc}D_2^*$ coupled systems with $J^P=3/2^+$, there are two $S$-wave components, the $\Xi_{cc}D_{1}|{}^4\mathbb{S}_{\frac{3}{2}}\rangle$ and $\Xi_{cc}D_{2}^{*}|{}^4\mathbb{S}_{\frac{3}{2}}\rangle$. In Table \ref{jg3}, although the bound eigenvalues for both of the isoscalar and isovector $\Xi_{cc}D_1/\Xi_{cc}D_2^*$ coupled systems with $J^P=3/2^+$ are similar to those from the single systems on the whole, the probabilities for the $\Xi_{cc}D_{2}^{*}|{}^4\mathbb{S}_{\frac{3}{2}}\rangle$ are obviously much larger. This indicates that the coupled channel effect is indeed helpful for the systems with very close masses and over one $S$-wave component.

Finally, we need to conclude again that the $\Xi_{cc}D_1/\Xi_{cc}D_2^*$ coupled systems with $I(J^P)=0(1/2^+,\,3/2^+)$ can be good triple-charm molecular candidates, and the dominant channel in the isoscalar $\Xi_{cc}D_1/\Xi_{cc}D_2^*$ coupled systems with $1/2^+$ is the $\Xi_{cc}D_1$ channel with almost 95\% probability. However, for the $I(J^P)=1(3/2^+)$ state, it is a mixture mainly composed by the $\Xi_{cc}D_1$ and $\Xi_{cc}D_2^*$ channel, and their probabilities are all over several tens of percents. Meanwhile, the isovector $\Xi_{cc}D_1/\Xi_{cc}D_2^*$ coupled systems with $J^P=(1/2^+,\,3/2^+)$ may be possible triple-charm molecular states.

\subsection{Other molecular pentaquarks}\label{sub3}

As a byproduct, we shall extend the obtained OBE effective potentials to the other systems composed by an $S$-wave doubly charmed baryon and an anticharmed meson in the $T$ doublet \cite{Klempt:2002ap}, i.e.,
\begin{equation}
V^{\Xi_{cc}{\bar T}\rightarrow\Xi_{cc}{\bar T}}({\bf{r}})=\sum_{E}(-1)^{G_E}V^{\Xi_{cc}T\rightarrow\Xi_{cc}T}_{E}({\bf{r}}),
\end{equation}
where $G_{E}$ stands for the $G$ parity for all the allowed exchanged mesons.

\renewcommand\tabcolsep{0.06cm}
\renewcommand{\arraystretch}{1.8}
\begin{table}[!htbp]
\caption{Bound state solutions for $\Xi_{cc} \bar D_{1}$ and $\Xi_{cc} \bar D_{2}^{*}$ systems. Notation $\times$ means no binding solutions. Here, the cutoff $\Lambda$, the binding energy $E$, and the root-mean-square radius $r_{rms}$ are in units of $ \rm {GeV}$, $\rm {MeV}$, and $\rm {fm}$, respectively.}\label{anti-charmed}
\begin{tabular}{cccccccc}\toprule[1.0pt]\midrule[1.0pt]
States&$\Lambda$ &$E$  &$r_{rms}$ &States&$\Lambda$ &$E$ &$r_{rms}$ \\\midrule[1.0pt]
$[\Xi_{cc}\bar D_{1}]_{J=1/2}^{I=0}$&1.16&$-0.48$ &4.11&$[\Xi_{cc} \bar D_{2}^{*}]_{J=3/2}^{I=0}$&1.15&$-0.37$ &4.64 \\
&1.21&$-3.82$&1.63&&1.23&$-6.53$&1.30 \\
&1.26&$-11.29$&1.03& &1.26&$-11.30$ &1.03\\
$[\Xi_{cc}\bar D_{1}]_{J=1/2}^{I=1}$&$\times$&$\times$&$\times$&$[\Xi_{cc}\bar D_{2}^{*}]_{J=3/2}^{I=1}$&$\times$&$\times$&$\times$\\
$[\Xi_{cc}\bar D_{1}]_{J=3/2}^{I=0}$&1.07&$-0.45$ &4.19&$[\Xi_{cc}\bar D_{2}^{*}]_{J=5/2}^{I=0}$&1.05&$-0.32$ &4.87\\
&1.21&$-5.91$&1.41&&1.18&$-4.84$&1.52 \\
&1.31&$-11.97$&1.09&&1.31&$-12.17$&1.09\\
$[\Xi_{cc}\bar D_{1}]_{J=3/2}^{I=1}$&2.25&$-0.55$ &3.27&$[\Xi_{cc}\bar D_{2}^{*}]_{J=5/2}^{I=1}$&2.04&$-1.54$ &1.90 \\
&2.26&$-2.49$&1.47&&2.05&$-3.92$ &1.18\\
&2.28&$-8.56$ &0.78&&2.07&$-10.49$ &0.72 \\
\bottomrule[1.0pt]\midrule[1.0pt]
\end{tabular}
\end{table}

In Table \ref{anti-charmed}, we present the bound state properties for the $\Xi_{cc} \bar D_{1}$ and $\Xi_{cc} \bar D_{2}^{*}$ systems. There exist six group bound states, the $\Xi_{cc}\bar D_{1}$ systems with $I(J^P)=0(1/2^+,\,3/2^+)$, $(1,3/2^+)$, and the $\Xi_{cc}\bar D_2^*$ states with $I(J^P)=0(3/2^+,\,5/2^+)$, $(1,5/2^+)$. Here, we find that the cutoff $\Lambda$ for the isovector $\Xi_{cc} \bar D_{1}$ and $\Xi_{cc} \bar D_{2}^{*}$ bound states are around 2.0 GeV, which is a little away from the typical value of the deuteron, 1.0 GeV. The remaining four isoscalar bound states can be prime loose molecular pentaquark candidates for their reasonable cutoff value, binding energy, and size.

\section{Conclusion and discussion}\label{sec5}

The hadronic molecular explanations of newly $X/Y/Z/P_c$ states inspire us to perform a systematic investigation of the possible triple-charm molecular states. If some of $X/Y/Z$ and $P_c$ states are really hidden-charm molecular tetraquarks and pentaquarks, respectively, one can naturally propose that there may exist possible triple-charm molecular pentaquarks, which are made up by a doubly charmed baryon and a charmed meson. In fact, we already studied the $\Xi_{cc}D/\Xi_{cc}D^*$ interactions \cite{Chen:2017jjn} and can predict several possible triple-charm molecular pentaquarks.

In this work, we further study the $S$-wave interactions between a doubly charmed baryon and a charmed meson in the $T$ doublet. According to the previous predictions in Ref. \cite{Chen:2017jjn}, at first, we qualitatively propose that interactions for the $\Xi_{cc}D_1$ systems with $I(J^P)=0(1/2^+),\,0(3/2^+)$ and $\Xi_{cc}D_2^*$ systems with $I(J^P)=0(3/2^+),\,0(5/2^+)$ can be strong enough to bind as bound states by using the HQSS analysis. Then, the OBE model is applied to the dynamical calculation, which is often adopted to study the interaction of heavy flavor hadrons after the observation of $X/Y/Z/P_c$ states. Since the coupling constants are not known, we need to calculate the coupling constants between the hadrons and light mesons with the quark model. In this work, we also introduce a cutoff parameter. According to the experience of the interaction between the nucleons and pion \cite{Tornqvist:1993ng,Tornqvist:1993vu}, a reasonable value of $\Lambda$ is taken around 1.0 GeV.

Finally, our numerical results show that the most promising pentaquark moleculars are the isoscalar $\Xi_{cc}D_1$ systems with $J^P=$ $(1/2^+,\,3/2^+)$ and the isoscalar $\Xi_{cc}D_2^*$ systems with $J^P=$ $(3/2^+,\,5/2^+)$, which is consistent with the results of the HQSS analysis. Although the cutoff value for the isovector $\Xi_{cc}D_1$ and $\Xi_{cc}D_2^*$ bound states is a little away from the empirical value $\Lambda$ around 1.0 GeV, we may predict that they may be possible triple-charm molecular candidates. Comparing the results whether considering the coupled channel effect or not, we find that (i) the coupled channel effect plays a negligible role in forming the loosely hadronic molecules for the $\Xi_{cc}D_1/\Xi_{cc}D_2^*$ state with $I(J^P)=0(1/2^+)$, where there is only one $S$-wave component, the $\Xi_{cc}D_1|{}^2\mathbb{S}_{\frac{1}{2}}\rangle$. (ii) For the $\Xi_{cc}D_1/\Xi_{cc}D_2^*$ state with $I(J^P)=0(3/2^+)$, there are two $S$-wave components, the $\Xi_{cc}D_1|{}^4\mathbb{S}_{\frac{3}{2}}\rangle$ and $\Xi_{cc}D_2^*|{}^4\mathbb{S}_{\frac{3}{2}}\rangle$; the coupled channel effect is very important.

As a byproduct, we extend our investigation to the $\Xi_{cc}\bar{D}_1$ and $\Xi_{cc}\bar{D}_2^*$ interactions, and there are several good possible molecular pentaquarks, the $\Xi_{cc}\bar D_{1}$ systems with $I(J^P)=0(1/2^+,\,3/2^+)$ and the $\Xi_{cc}\bar D_2^*$ states with $I(J^P)=0(3/2^+,\,5/2^+)$.

\section*{ACKNOWLEDGMENTS}
This project is supported by the China National Funds for Distinguished Young Scientists under Grants No. 11825503. This project is also supported by the National Natural Science Foundation of China under Grants No. 11705072.

\appendix*
\section{Coupling constants}\label{app01}

The concrete expressions for interactions between $T $doublet charmed mesons and light mesons are written as
\begin{itemize}
  \item scalar meson $\sigma$ exchange:
\begin{eqnarray}
&&\langle t \rangle^{H}=ig_\sigma^{\prime\prime},\nonumber\\
&&\langle t \rangle^{Q}=ig_{\sigma qq}<D_{2}^{*}\mid D_{2}^{*}>=ig_{\sigma qq}.
\end{eqnarray}
  \item pseudoscalar meson $\pi$ exchange:
\begin{eqnarray}
\langle t \rangle^{H}&=&-\frac{k}{f_\pi}q^{3},\nonumber\\
\langle t \rangle^{Q}&=&\frac{g_{qq\pi}}{\sqrt{2}M_q}q^{3}<D_{2}^{*}\uparrow\mid\sum_{x}\sigma_{x}^3\tau_{x}^3\mid D_{2}^{*}\uparrow>=-\frac{g_{qq\pi}}{\sqrt{2}M_q}q^{3}.\nonumber\\
\end{eqnarray}
  \item vector meson $\rho$ exchange:
\begin{eqnarray}
\langle t \rangle^{H}&=&\frac{i}{\sqrt{2}}\beta^{\prime \prime}g_{V}\rho_3^{0}-\sqrt{2}\lambda^{\prime\prime} g_{V}\epsilon_{3ij}\rho_{3}^{i}q^{j},\nonumber\\
\langle t \rangle^{Q}&=&\sqrt{2}ig_{qq\rho}\rho_{3}^{0}<D_{2}^{*} \mid \sum_{x}\tau_{x}^{3} \mid D_{2}^{*}>+\Big(\frac{g_{qq\rho}}{\sqrt{2}M_q}\nonumber\\
&&+\frac{f_{qq\rho}}{\sqrt{2}M_N}\Big)\epsilon_{3ij}\rho_{3}^{i}q^{j}<D_{2}^{*} \uparrow \mid \sum_{x} \sigma_{x}^{3}\tau_{x}^{3} \mid D_{2}^{*}\uparrow>\nonumber\\
&=&-\sqrt{2}ig_{qq\rho}\rho_{3}^{0}-\Big(\frac{g_{qq\rho}}{\sqrt{2}M_q}+\frac{f_{qq\rho}}{\sqrt{2}M_N}\Big)\epsilon_{3ij}\rho_{3}^{i}q^{j}.
\end{eqnarray}
\end{itemize}
Here, $q^i$ is the $i$ component of the exchanged boson momentum, notation $\uparrow$ means the third component of the spin is $+2$. The superscripts $H$ and $Q$ stand for the interactions in the hadronic and quark level, respectively. After comparing these interactions in hadronic level and quark level, one can obtain several coupling constants relations. Using the same method, coupling constants for $T$ doublet charmed mesons and light meson interactions can be related to those from nucleon-nucleon interactions, i.e.,
\begin{eqnarray}
&&g_\sigma^{\prime\prime}=\frac{1}{3}g_{\sigma NN},~~\frac{k}{f_\pi}=\frac{3}{5\sqrt{2}}\frac{g_{\pi NN}}{M_N},\nonumber\\
&&\beta^{\prime\prime}g_{V}=-2g_{\rho NN},~~\lambda^{\prime\prime} g_{V}=\frac{3}{10}\frac{g_{\rho NN}+f_{\rho NN}}{M_N}.
\end{eqnarray}

\end{document}